\documentclass[prx,twocolumn,groupedaddress]{revtex4}
\usepackage{graphicx,color}
\usepackage{amssymb}   % for math
\usepackage{amsmath}
\usepackage{epstopdf}
\usepackage{natbib}
\usepackage{hyperref}
\usepackage{bm}
\usepackage{color}
\usepackage{braket}

\begin{document}
\title{Highly tunable nonlinear Hall effects induced by spin-orbit couplings \\
in strained polar transition-metal dichalcogenides}
\author{Benjamin T. Zhou$^{1}$} \thanks{Correspondence author: bentzhou@ust.hk\\ These authors contributed equally to this work.}
\author{Cheng-Ping Zhang$^{1}$} \thanks{These authors contributed equally to this work.}
\author{K. T. Law$^{1}$} \thanks{Correspondence author: phlaw@ust.hk}

\affiliation{$^{1}$Department of Physics, Hong Kong University of Science and Technology, Clear Water Bay, Hong Kong, China} 

\begin{abstract}
Recently, signatures of nonlinear Hall effects induced by Berry curvature dipoles have been found in atomically thin 1T'/T$_d$-WTe$_2$. In this work, we show that in strained polar transition-metal dichalcogenides(TMDs) with 2H-structures, Berry curvature dipoles created by spin degrees of freedom lead to strong nonlinear Hall effects. Under easily accessible uniaxial strain of order $\sim 0.2\%$, strong nonlinear Hall signals, characterized by Berry curvature dipole in the order of $\sim 1 {\AA}$, arise in electron-doped polar TMDs such as MoSSe, which is easily detectable experimentally. Moreover, the magnitude and sign of the nonlinear Hall current can be easily tuned by electric gating and strain. These properties can be used to distinguish nonlinear Hall effects from classical mechanisms such as ratchet effects. Importantly, our system provides a potential scheme for building electrically switchable energy harvesting rectifiers. 
\end{abstract}

\maketitle

\section{\bf{Introduction}}

The study of Hall effects has been one of the central topics in condensed matter physics\cite{Niu, Nagaosa}. Within the linear response regime, Hall effect arises only when time-reversal symmetry is broken\cite{Landau, Culcer}. Recently, however, it was proposed by Sodemann and Fu\cite{Sodemann} that Hall effects can occur in a wide class of time-reversal-invariant materials with broken spatial inversion symmetry. In such systems, the total Berry flux over the equilibrium distribution is zero due to time-reversal symmetry\cite{Niu}, while Berry curvatures can emerge locally in the Brillouin zone, with counter-propagting charge carriers having different Berry curvatures. Under an applied electric field, the current-carrying state maintains an imbalance between counter-propagting movers, which results in nonzero Berry curvature flux under proper symmetry conditions\cite{Lee}. This leads to anomalous Hall currents which establish a Hall voltage in the steady state. As the electric field plays both roles of driving the system out of equilibrium and inducing anomalous velocities, the Hall current scales quadratically with the voltage bias. This special type of Hall effect is thus referred to as the nonlinear Hall effect(NHE). Due to the nonlinear current response, the NHE can convert oscillating electric fields into DC currents, a process known as rectification, which have potential applications for next-generation wireless and energy harvesting devices\cite{Isobe}.

Nonlinear Hall response is characterized by the first-order moment of Berry curvatures over occupied states\cite{Sodemann, Guinea, Spivak}, called the Berry curvature dipole. In 2D systems, the Berry curvature dipole transforms as a pseudo-vector, thus the maximum symmetry allowed for a nonzero moment is a single mirror symmetry (mirror plane perpendicular to the 2D plane). Interestingly, atomically thin transition-metal dichalcogenides(TMDs) with 1T'/T$_d$-structure, such as MoTe$_2$ and WTe$_2$, respect a single in-plane mirror symmetry\cite{Car}, and nonzero Berry curvature dipoles are proposed to exist in these materials\cite{You, Yan}. Remarkably, two recent experiments have independently observed signatures of Hall effects in bilayer\cite{Pablo} and multi-layer WTe$_2$\cite{Fai} in the absence of magnetic fields. Importantly, a quadratic scaling relation was found between the transverse voltage and the applied source-drain bias. While the observations are consistent with NHEs induced by Berry curvature dipoles\cite{Haizhou}, due to the relatively weak gate-dependence of Berry curvature dipoles in WTe$_2$\cite{Pablo}, it remains experimentally challenging to directly rule out alternative trivial interpretations such as electron ratchet effects\cite{Sodemann, Fai, Ganichev}. 

Besides 1T'/T$_d$-TMDs, it is known that TMDs with the usual 2H-structures also possess nontrivial Berry curvatures due to intrinsically broken inversion in orbital degrees of freedom\cite{Xiao, Fai1}. However, the three-fold($C_3$) symmetry in 2H-TMDs forces the Berry curvature dipole to vanish\cite{Sodemann}. Under applied strains that break the $C_3$-symmetry, nonzero Berry curvature dipoles can arise in 2H-TMDs\cite{Sodemann, Lee, You}. Unfortunately, the dipole moment in strained conventional 2H-TMDs is shown to be weak ($\sim 0.01 {\AA}$)\cite{Sodemann, You}, mainly due to the weak Berry curvatures generated by the huge Dirac mass ($\sim 1$ eV) in the orbital degrees of freedom\cite{Xiao}. 

Lately, it was proposed that in gated\cite{Ye1} or polar 2H-TMDs\cite{Wan, Cheng}, combined effects of Rashba and Ising spin-orbit couplings(SOCs) result in a different type of Berry curvature in spin degrees of freedom\cite{Benjamin, Taguchi}. Importantly, in 2H-TMDs the SOC-induced effect was found to be strong, which dominates over the conventional orbital effect and significantly changes the Berry curvatures in 2H-TMDs. However, the role of this SOC-induced Berry curvatures in creating nonlinear Hall effects in 2H-TMDs remains unknown.

\begin{figure}
\centering
\includegraphics[width=3.5in]{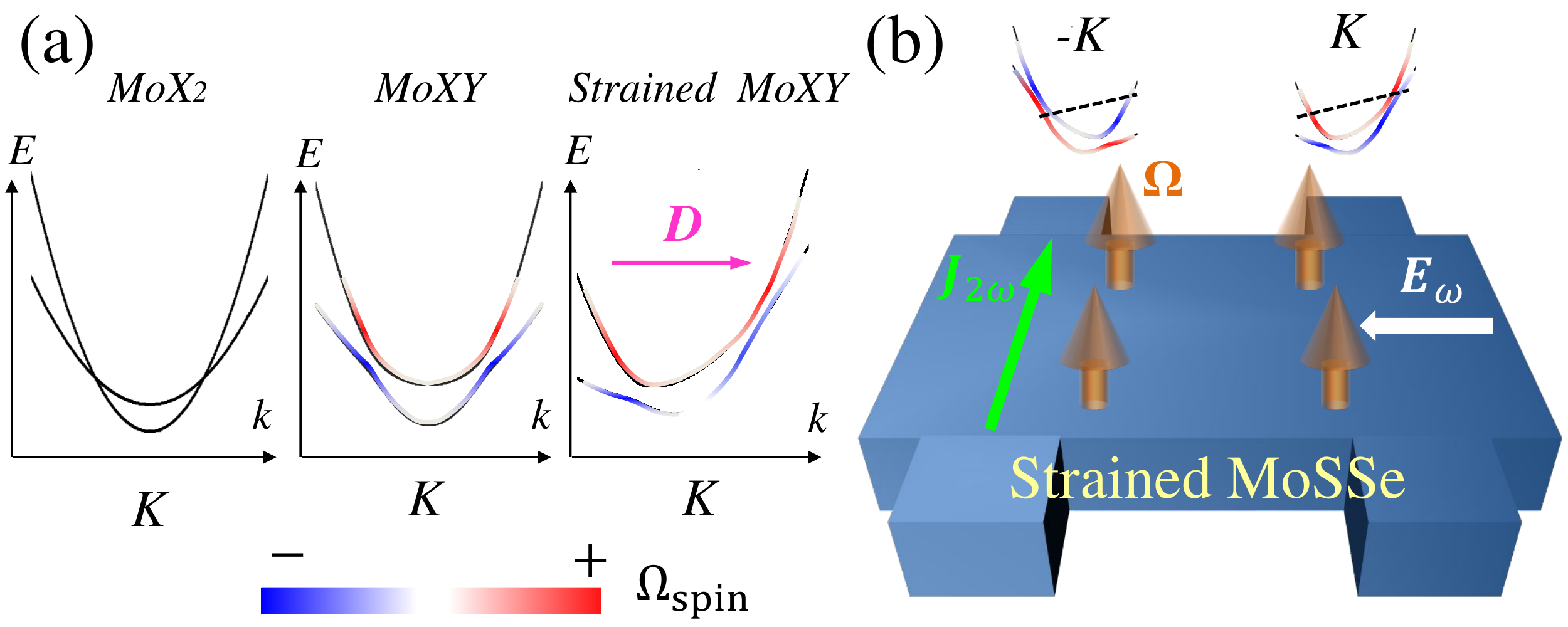}
\caption{Schematics for Berry curvature dipole in strained moly-based transition-metal dichalcogenides(TMDs). (a) $\Omega_{spin}$ near the conduction band minimum(CBM) at the $+K$-point in 2H-TMDs. Left panel: in MoX$_2$, spin-up and spin-down bands cross at finite momentum. $\Omega_{spin}$ is zero due to the absence of Rashba SOCs. Middle panel: in MoXY, Rashba SOCs cause anti-crossings within two spin-subbands, where hot spots of $\Omega_{spin}$ emerge at the same energy for left-movers and right-movers. Right panel: with $u_{xx} \neq 0$, anti-crossings for left-movers and right-movers are separated, creating nonzero Berry curvature dipole $\textbf{D}$ in MoXY. (b) Under electric field $\textbf{E}_{\omega}$ with frequency $\omega$, the non-equilibrium state in strained MoSSe gains net Berry curvature flux (denoted by the orange arrows), which combines with $\textbf{E}_{\omega}$ to generate nonlinear Hall current signified by second-harmonic component $\textbf{J}_{2\omega}$.}
\label{FIG01}
\end{figure}

In this work, we show that the large Berry curvatures induced by SOCs lead to strong and gate-tunable nonlinear Hall effects(NHEs) in moly-based polar 2H-TMDs(Fig.\ref{FIG01}) such as MoSSe\cite{Ang-Yu, Jing}. With easily accessible low carrier density and weak uniaxial strain of order $\sim 0.2\%$, pronounced nonlinear Hall signals arise in polar TMDs characterized by Berry curvature dipoles on the order of $\sim 1 {\AA}$(Fig.\ref{FIG03}), comparable to the optimal values observed recently in 1T$_d$-WTe$_2$\cite{Pablo, Fai}. 

Importantly, the magnitude and sign of nonlinear Hall signals in strained polar TMDs change dramatically upon gating the Fermi level $\sim 10-20$ meV away from the conduction band minimum(Fig.\ref{FIG02}(c)). Therefore, nonlinear Hall effects in strained polar TMDs generally exhibit a stronger gate-dependence than 1T'/T$_d$-WTe$_2$\cite{Pablo}, which can be easily detected in Hall measurements with moderate gating. We further point out that the gate-sensitive NHE in strained polar TMDs provides a promising scheme for realising electrically switchable rectifiers for wireless energy harvesting devices. The highly gate-tunable NHE in strained polar TMDs also serves as a distinctive and accessible experimental signature of Berry curvature dipoles, which distinguishes itself from nonlinear effects due to trivial classical mechanisms\cite{Sodemann, Fai, Ganichev}. 

\section{\bf Results}
\subsection{\bf Effective model Hamiltonian of strained MoSSe}

Throughout this work, we consider MoSSe as a specific example of polar TMDs, which has been successfully fabricated in recent experiments\cite{Ang-Yu, Jing}. However, our prediction generally applies to the whole class of moly-based polar TMDs MoXY, (X $\neq$ Y)\cite{Wan, Cheng}. 

To distinguish the two types of Berry curvatures in 2H-TMDs originating from orbital/spin degrees of freedom, we use the notations $\Omega_{orb}$/$\Omega_{spin}$ to denote the conventional Berry curvatures/the SOC-induced Berry curvatures.

To describe the essential mechanism behind the emergence of Berry curvature dipole, we first construct an effective Hamiltonian for $n$-type(electron-doped) strained polar TMDs. The crystal structure of a generic polar TMD is almost the same as a usual 2H-TMD except that the triangularly arranged transition-metal(M) atoms are sandwiched by two different layers of chalcogen atoms\cite{Ang-Yu, Jing, Wan, Cheng}. Thus, the out-of-plane mirror symmetry (mirror plane parallel to the 2D plane of M-atoms), which is generally respected by usual 2H-TMDs, is intrinsically broken. The resultant symmetry group of MoSSe is the product group of the $C_{3v}$ point group and time-reversal symmetry $\mathcal{T}$.

Near the conduction band minimum(CBM) at the $K$-points, electrons in polar TMDs originate predominantly from the $d_{z^2}$-orbitals of the M-atoms\cite{Liu}. Under the basis formed by spins of $d_{z^2}$-electrons, the unstrained effective Hamiltonian of MoSSe can be written as\cite{Benjamin, Taguchi, Noah}:
\begin{eqnarray}\label{eq:01}
H_0 &=& \xi_{\bm{k}} \sigma_0 + \alpha_{so} (k_y \sigma_x - k_x \sigma_y) + \epsilon \beta_{so}(\bm{k}) \sigma_z .
\end{eqnarray}
Here, $\sigma_{i}, i = 0,x,y,z$ denotes the usual Pauli matrices acting on the spin degrees of freedom. $\xi_{\bm{k}} =\frac{|\bm{k}|^2}{2m^{\ast}} - \mu$ denotes the kinetic energy term, $m^{\ast}$ is the effective mass of the electron band, $\mu$ is the chemical potential, $\epsilon=\pm$ is the valley index. 

The $\beta_{so}(\bm{k})$-term refers to the Ising SOC which originates from the atomic spin-orbit coupling as well as the breaking of an in-plane mirror symmetry\cite{Yuan, Zhu, Kormanyos, Kosmider}. In previous studies on 2H-TMDs, the Ising SOC was usually treated as a constant near the $K$-points\cite{Benjamin, Taguchi, Noah}. However, in realistic band structures of conventional moly-based 2H-TMDs, spin-up and spin-down bands cross at finite momentum $k_0$ (left panel of Fig.\ref{FIG01}(a))\cite{Liu}. This indicates a sign change in the Ising SOC term at the crossing, which can be accounted by quadratic corrections in $\bm{k}$: $\beta_{so}(\bm{k}) = \beta_0 + \beta_1 k^2$, with $sgn(\beta_0) sgn(\beta_1) <0$. The crossing for spin-up and spin-down bands occurs at $k_0 = \pm \sqrt{-\beta_0/\beta_1}$. The $\alpha_{so}$-term is known as the Rashba SOC\cite{Rashba}, which arises in MoSSe due to intrinsically broken out-of-plane mirror symmetry.

Interestingly, per each $K$-valley the coexistence of Ising and Rashba SOCs results in a modified massive Dirac Hamiltonian $H_0 = \xi_{\bm{k}} \sigma_0 + \bm{d}_{\epsilon}(\bm{k})\cdot \bm{\sigma}$, which is reminiscient of the well-known BHZ model for a 2D topological insulator\cite{BHZ} with $\bm{d}_{\epsilon}(\bm{k}) = [\alpha_{so} k_y, -\alpha_{so} k_x, \epsilon (\beta_0 + \beta_1 k^2) ]$ except that the kinetic $\xi_{\bm{k}}$-term bends the electron bands up, with two non-degenerate  spin-subbands(Fig.\ref{FIG01}(a)). The energy spectra for the upper/lower spin subbands are given by: $E_{\pm} (\bm{k} + \epsilon \bm{K}) = \xi_{\bm{k}} \pm \epsilon |\bm{d}(\bm{k})|$ where $|\bm{d}(\bm{k})| = \sqrt{\alpha_{so}^2k^2 + (\beta_0 + \beta_1 k^2)^2}$. The Berry curvatures of the two subbands are given by\cite{Qi, Volovik}: 
\begin{eqnarray}\label{eq:02}
\Omega_{spin, \pm}(\bm{k} + \epsilon \bm{K}) &=& \mp \frac{1}{2} \hat{\bm{d}}_{\epsilon}(\bm{k}) \cdot ( \frac{\partial \hat{\bm{d}}_{\epsilon} }{\partial k_x} \times \frac{\partial \hat{\bm{d}}_{\epsilon} }{\partial k_y}) \\\nonumber
&=& \mp \frac{1}{2} \epsilon \frac{ \alpha_{so}^2 (\beta_0 - \beta_1 k^2)}{|\bm{d}(\bm{k})|^{3}}.
\end{eqnarray}
Note that the nontrivial $\Omega_{spin, \pm}(\bm{k})$ requires the presence of both Ising and Rashba SOCs. In the absence of Rashba SOCs, the upper and lower subbands consist of decoupled spin-up and spin-down states and no $\Omega_{spin}$ can be generated (left panel of Fig.\ref{FIG01}(a)). In moly-based polar TMDs, the Rashba SOC hybridizes the spin-up and spin-down bands and causes an anti-crossing within the spin subbands (middle panel of Fig.\ref{FIG01}(a)). In particular, hot spots of $\Omega_{spin}$ emerge in the vicinity of the crossing points $k_0$, with their signs being opposite in upper and lower subbands. Given realistic parameters, $\Omega_{spin}$ near its hot spots has a magnitude $|\Omega_{spin}| \gg 100 {\AA}^2$ (see Appendix A for details).

Despite the large $\Omega_{spin}$, the Berry curvature dipole remains zero in unstrained MoSSe due to the three-fold($C_3$) symmetry\cite{Sodemann}. Physically, the Berry curvature dipole measures the gain in total Berry curvature flux in the current-carrying state\cite{Lee}. When $C_3$-symmetry is present, the Berry flux from left-movers is always equal to that from right-movers on the Fermi surface (middle panel of Fig.\ref{FIG01}(a)). Thus, the imbalance between left-movers and right-movers established by a source-drain bias leads to no gain in total Berry flux up to the lowest order correction.

To break the $C_3$-symmetry, one feasible way is to introduce uniaxial strains\cite{Sodemann, Lee, You}. Following the scheme developed in the recent work\cite{Shiang}, effects of strains in 2D TMDs can be modelled by classifying the strain-field tensor $\overleftrightarrow{\bm{u}}_{ij} = (\partial_{i} u_{j} + \partial_{j} u_{i})/2, (i,j=x,y$) according to the irreducible representations of the $C_{3v}$ point group of polar TMDs. In total, $\overleftrightarrow{\bm{u}}$ has three effective independent components: (i) the trace scalar $u_0 \equiv (u_{xx} + u_{yy})$; and (ii) the doublet $\{u_{1}, u_{2}\} \equiv \{2u_{xy}, u_{xx}-u_{yy}\}$ that transforms as a polar vector. Details of the symmetry properties of $u_0, u_1, u_2$ are presented in Appendix A. 

To capture the essential physics, we consider strain effects on spin-independent terms only, which has no contributions to Berry curvatures. This approximation is based on the observation that the coupling strength between the spin-independent terms and strain-field  $\overleftrightarrow{\bm{u}}$ have an energy scale $\sim 1$ eV\cite{Shiang}, which is far greater than spin-orbit couplings on the scale of a few to tens of meVs. Up to linear order terms in $\bm{k}$, the effective strained Hamiltonian compatible with the $C_{3v} \otimes \mathcal{T} $ symmetry group is given by:
\begin{eqnarray}\label{eq:03}
H_{strain} &=&  [\gamma u_{0} + \epsilon \delta (u_{1} k_y - u_{2}k_x ) ]\sigma_0 ,\\\nonumber
H_{eff} &=& H_0 + H_{strain},
\end{eqnarray}
where $H_{eff} $ is the total effective Hamiltonian, and $\gamma$, $\delta$ are effective strained parameters. Considering uniaxial strains in the $x$-direction: $u_{xx} \neq 0$ and $u_{xy} = u_{yy} = 0$, the strained energy spectra are given by: $E'_{\pm} (\bm{k}) =u_{xx} (\gamma  - \epsilon \delta k_x ) + \xi_{\bm{k}} \pm |\bm{d}_{\epsilon}(\bm{k})|$. Clearly, the $\delta$-term breaks the $C_3$-symmetry by shifting the band minimum along the $k_x$ direction\cite{Sodemann}. As a result, the two pairs of Berry curvature hot spots associated with the left-movers and right-movers are energetically separated(Fig.\ref{FIG01}(a), right panel). For Fermi level close to one of these separated hot spots, an applied bias in the $x$-direction creates an imbalance between left-movers and right-movers, and the system acquires nonzero out-of-plane Berry curvature flux in the current-carrying state(orange arrows in Fig.\ref{FIG01}(b)). This current-induced Berry flux then combines again with the applied field to generate currents in transverse $y$-direction as shown schematically in Fig.\ref{FIG01}(b).

\subsection{\bf{Large and gate-tunable Berry curvature dipoles in strained MoSSe}}

Having established how $\Omega_{spin}$ combines with uniaxial strains to create Berry curvature dipoles, we now go beyond the effective two band model and study the nonlinear Hall effect in strained MoSSe under realistic situations. Using a six-orbital tight-binding model for strained TMDs\cite{Liu, Shiang}, we take both $\Omega_{spin}$ and $\Omega_{orb}$ into account and study the Berry curvature as well as its dipole moment in electron-doped strained MoSSe. As we are about to show, the dominance of $\Omega_{spin}$ over $\Omega_{orb}$, together with uniaxial strains, leads to strong and highly gate-tunable nonlinear Hall effects in strained MoSSe. Details of the tight-binding Hamiltonian are presented in Appendix B-C.

As we discussed in the last section, the Berry curvature dipole measures the lowest-order correction in total Berry curvature flux in the non-equilibrium state. This physical meaning is revealed by its formal expression \cite{Sodemann, Pablo}:
\begin{eqnarray}\label{eq:BCDipole}
D_{\alpha} &=& -\sum_{n} \int \frac{d^2 \bm{k}}{(2\pi)^2} (\partial_{k_{\alpha}} f_n) \Omega_{n}(\bm{k})\\\nonumber
&=& \sum_{n} \int \frac{d^2 \bm{k}}{(2\pi)^2} v_{n, \alpha}(\bm{k}) \delta_{F}(E_{n}-E_{F}) \Omega_{n}(\bm{k}).
\end{eqnarray}
Here, $D_{\alpha}$ denotes the $\alpha$-component of the Berry curvature dipole $\bm{D}$, with $\alpha=x,y$. $f_n, \Omega_n$ refer to the equilibrium distribution function and Berry curvature of bands indexed by $n$. $v_{n, \alpha} = \partial E_{n}/\partial k_{\alpha}$ is the band velocity and $\delta_{F}(E-E_{F}) = -\partial f/ \partial E = [ 4k_B T \cosh^2(\frac{E-E_F}{2k_B T}) ]^{-1}$ mimics a delta-function with its maximum value $\delta_{F}^{max} = 1/4k_B T$ centered at $E_{F}$.

\begin{figure}
\centering
\includegraphics[width=3.5in]{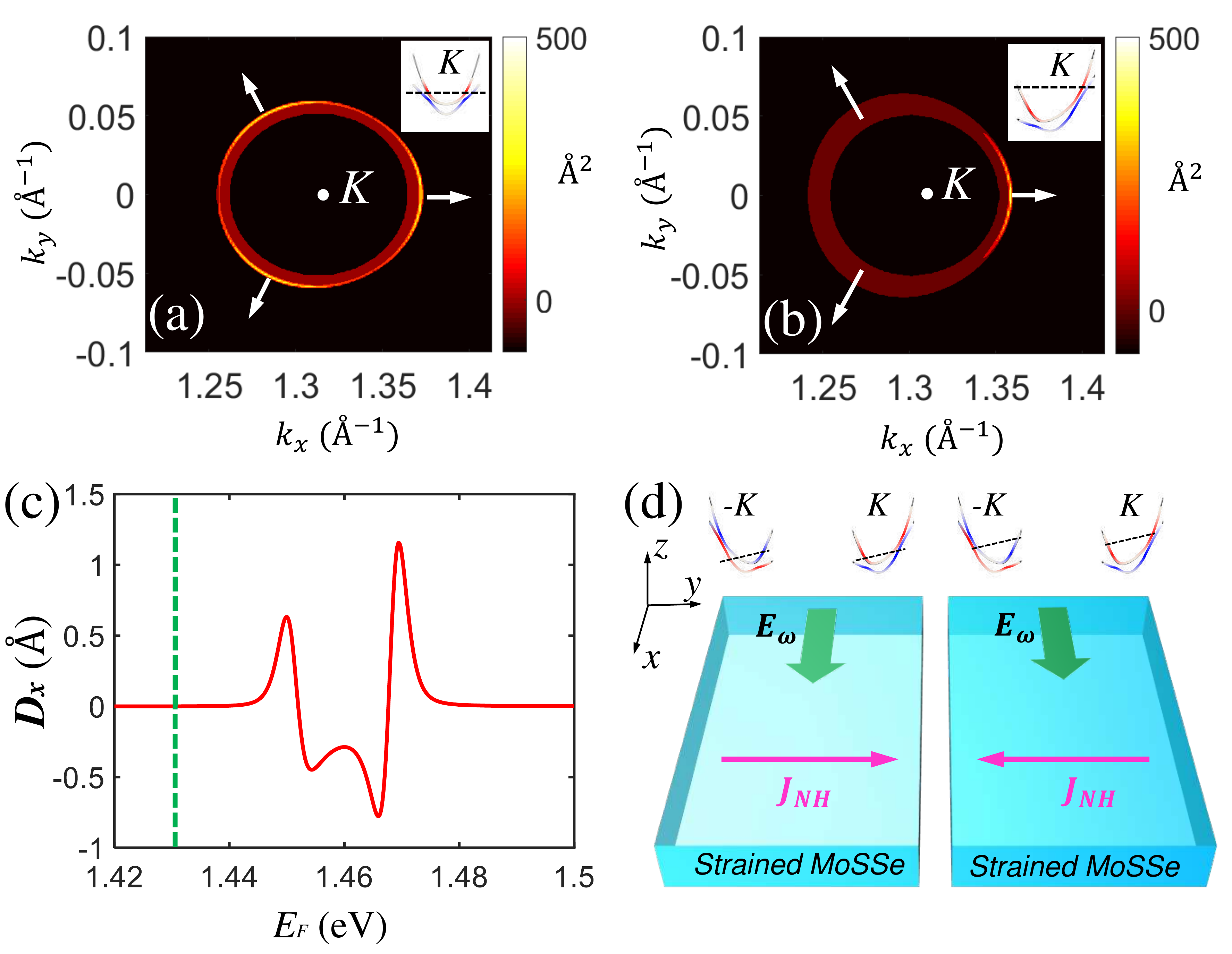}
\caption{Realistic Berry curvature and its dipole moment in MoSSe. Berry curvature profiles on the Fermi surface contour in (a) unstrained MoSSe, and (b) strained MoSSe. Contributions from both spin subbands are included. Berry curvatures are weighted by the normalized delta function $\delta_{F}/\delta^{max}_{F}$, with temperature $T = 10K$. In-sets in (a)-(b): schematics for locations of $E_F$ in each case. Colors of the bands indicate the Berry curvature values. (c) Berry curvature dipole $D_x$ versus Fermi energy $E_{F}$ in strained MoSSe (red solid line) calculated from realistic tight-binding model under uniaxial strain $u_{xx} = 4\%$. The green dashed line indicates the location of band minimum. (d) Schematics for strong gate-dependence of nonlinear Hall effects in strained MoSSe. By gating $E_{F}$ across the band anti-crossing associated with the right-movers in the $+K$-valley, the nonlinear Hall current $\bm{J}_{NH}$ switches sign. }
\label{FIG02}
\end{figure}

In general, the Berry curvature dipole in a polar TMD has contributions from both $K$ and $-K$ valleys. Due to time-reversal symmetry, $v_{n, \alpha}(\bm{k} + \bm{K}) = -v_{n, \alpha}(-\bm{k} - \bm{K})$, $\Omega_{n}(\bm{k} + \bm{K}) = -\Omega_{n}(-\bm{k}-\bm{K})$. Thus, contributions from the two $K$ valleys are equal, which allows us to consider the $+K$-valley only for the simplicity of our following discussions.

As explained previously, Berry curvature dipoles vanish in MoSSe in the absence of strains. Since the Berry curvature dipole is a Fermi liquid property\cite{Sodemann}, this symmetry property can be explicitly revealed by the Berry curvature profile on the Fermi surface contour of an unstrained MoSSe. Without loss of generality, we consider $E_F$ lying slightly above the Berry curvature hot spot in the upper band (in-set of Fig.\ref{FIG02}(a)). Apparently, due to the $C_3$-symmetry, Berry curvatures of three-fold-related momentum states $\{ \bm{k}, C_3\bm{k}, C^2_3\bm{k}\}$ must satisfy: $\Omega_{n}(\bm{k}) = \Omega_{n}(C_3\bm{k}) = \Omega_{n}(C^2_3\bm{k})$ (Fig.\ref{FIG02}(a)), and their band velocities (indicated by arrows in Fig.\ref{FIG02}(a)) sum to zero: $\sum_{j=1}^{3} v_{n,\alpha}(C^{j-1}_3 \bm{k}) = 0$. These symmetry constraints force contributions from left-movers and right-movers to cancel each other, leading to vanishing dipole moments in Eq.\ref{eq:BCDipole}. 

However, due to the fact that the total Berry curvature $\Omega_{tot}$ near the band anti-crossing points is approximately a sum of $\Omega_{spin}$ and $\Omega_{orb}$(see Appendix D for details), the dominance of $\Omega_{spin}$ over $\Omega_{orb}$ implies that the behavior of $\Omega_{tot}$ is essentially governed by $\Omega_{spin}$. Therefore, $\Omega_{tot}$ in both upper and lower subbands exhibit similar nonuniform momentum-space profiles as $\Omega_{spin}$ (see Appendix D for details), with hot spots emerging near the anti-crossing points (shown schematically in the in-set of Fig.\ref{FIG02}(a)). 

Importantly, $\Omega_{tot}$ in the upper and lower subbands have opposite signs\cite{Benjamin}(indicated by the red/blue colors in the in-sets of Fig.\ref{FIG02}(a)-(b)). While this generally leads to partial cancellation within the two subbands, it is important to note that for Fermi level located slightly above(below) the anti-crossing points, the Fermi momentum of the upper(lower) subband is closer to the hot spots, thus its Berry curvature contribution dominates over the other subband at the Fermi energy, leading to a large net Berry flux per each $K$-valley on the Fermi surface contour. For instance, for Fermi level slightly above the band anti-crossing, the upper subband dominates, with the overall sign of the Berry curvature being positive around $K$(Fig.\ref{FIG02}(a)).

Under uniaxial strains, the anti-crossings associated with the left-movers and right-movers are energetically separated. Thus, for Fermi levels located slightly above or below one of the separated anti-crossing points, there generally exists a huge difference between Berry curvature contributions from left-movers and right-movers. To be specific, we plot the Berry curvature profile on the Fermi surface contour of strained MoSSe under $u_{xx} = 4\%$, with $E_F$ lying slightly higher than the anti-crossing point associated with the right-movers (in-set of Fig.\ref{FIG02}(b)). Apparently, contributions from the right-movers far exceed those from left-movers, with the overall sign being positive due to the dominance of the upper band (Fig.\ref{FIG02}(b)). In this case, the system gains a large amount of Berry flux as the left-movers are pumped by the voltage bias to the right or vice versa, which signifies a large Berry curvature dipole. 

Moreover, since the upper/lower subband dominates for Fermi levels above/below the anti-crossing, the net Berry curvature on the Fermi surface changes sign as the Fermi level is gated across these anti-crossing points, indicating a sign switch in the Berry curvature dipole (Eq.\ref{eq:BCDipole}). The Berry curvature dipole $D_x$ as a function of $E_F$ under $u_{xx} = 4\%$ is plotted in Fig.\ref{FIG02}(c). At the band minimum(indicated by green dashed line in Fig.\ref{FIG02}(c)), the Berry curvature dipole is zero due to the vanishing band velocity. As $E_F$ increases, the anti-crossing associated with the left-movers is first accessed. Evidently, $D_x$ changes from positive to negative values as $E_F$ goes across the anti-crossing. This gives rise to a peak followed by a dip as shown in the $D_x-E_F$ curve ($E_F$ in the range $1.44 \sim 1.46$ eV in Fig.\ref{FIG02}(c)).

By further raising the Fermi level, one reaches the anti-crossing associated with the right-movers. However, since the band velocities of right-movers and left-movers are opposite to each other, $D_x$ changes sign in an opposite manner according to Eq.\ref{eq:BCDipole}, \textit{i.e.}, from negative to positive values as $E_F$ sweeps across the anti-crossing. This leads to a dip followed by a peak in the $D_x-E_F$ curve ($E_F$ in the range $1.46 \sim 1.48$ eV in Fig.\ref{FIG02}(c)).

Consider a strained MoSSe with coordinates defined in Fig.\ref{FIG02}(d)), an oscillating electric field with frequency $\omega$: $\bm{E}_{\omega}(t) = \text{Re} \{ \mathcal{E}_x e^{i\omega t} \} \hat{x}$ in the $x$-direction can drive a second-harmonic transverse current along $y$-direction in strained MoSSe(Fig.\ref{FIG01}(b)), with the current amplitude given by\cite{Sodemann}:
\begin{eqnarray}
j^{2\omega}_y = -\frac{e^3 \tau \mathcal{E}^2_x }{2 (1+i\omega \tau)} D_x,
\end{eqnarray}
where $\tau$ is the relaxation time. As $j^{2\omega}_y$ is proportional to $D_x$, the nonlinear Hall current in MoSSe also changes sign upon gating the Fermi level across the anti-crossing. This sign switch occurs within a narrow Fermi level range $\Delta E_{F} \sim 10$ meV(Fig.\ref{FIG02}(d)) and thus can be easily controlled by moderate gating. The second-harmonic Hall current $j^{2\omega}_y$ establishes an AC Hall voltage with frequency $2\omega$, which can be readily detected by usual Hall bar geometry as in recent experiments on bilayer and multilayer WTe$_2$\cite{Pablo, Fai}. 

\begin{figure}
\centering
\includegraphics[width=3.5in]{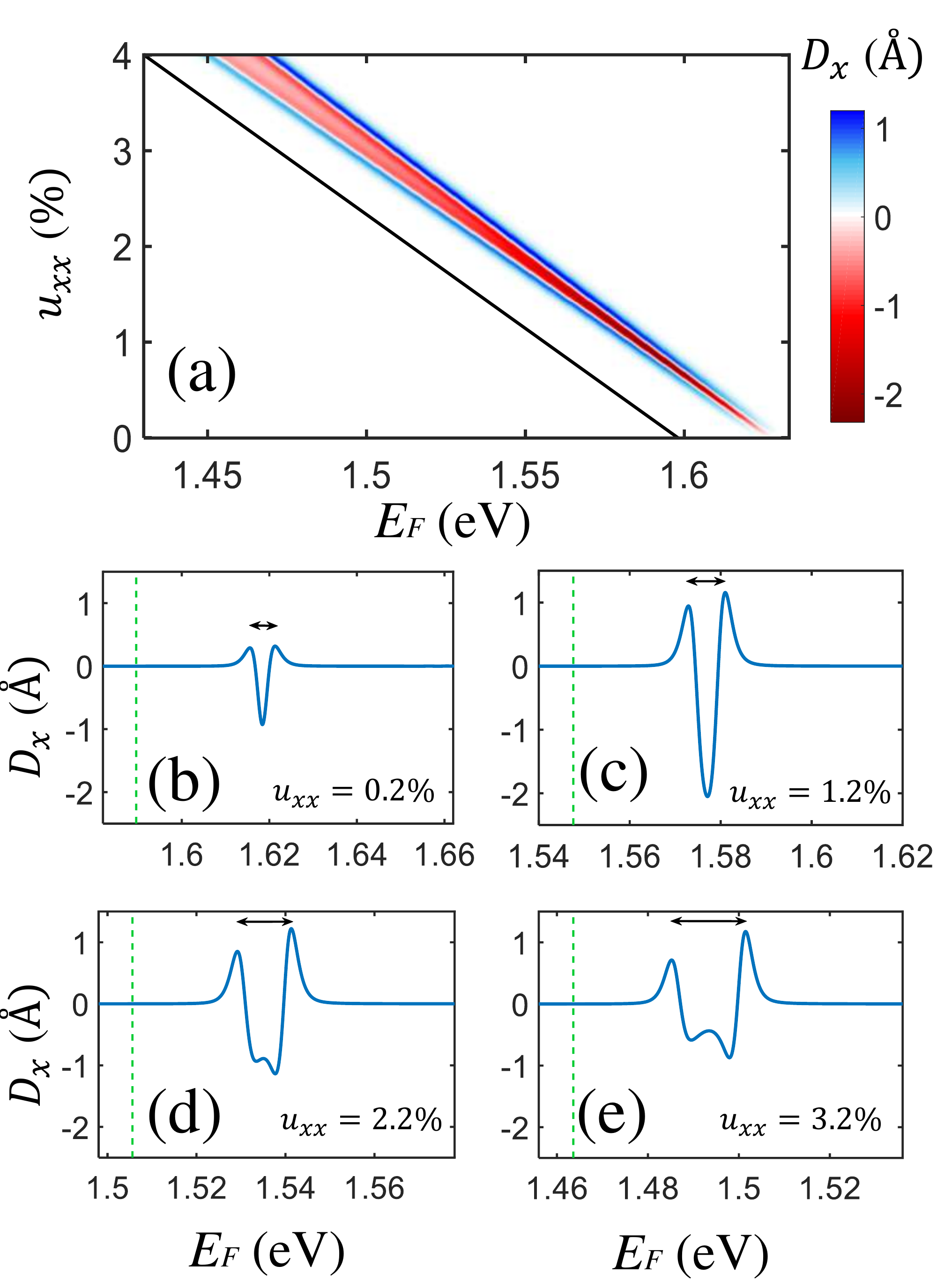}
\caption{Evolution of nonlinear Hall signals induced by Berry curvature dipoles in MoSSe under strains and gating. (a) Strain-gate map of Berry dipole $D_x$ in electron-doped MoSSe. Black solid line indicates location of conduction band minimum(CBM). Note that the energy offsets in CBM upon increasing $u_{xx}$ is due to on-site energy corrections induced by strains (see Appendix C). (b)-(e) Gate-dependence of $D_x$ with $u_{xx}=0.2\%, 1.2\%, 2.2\%, 3.2\%$. Green dashed lines correspond to locations of CBM in each case. The black double arrows indicate the energy separation between anti-crossings associated with the left-movers and the right-movers near the $K$-valley. }
\label{FIG03}
\end{figure}

\subsection{\bf{Strain-gate map of nonlinear Hall response in MoSSe}}

In this section, we systematically demonstrate how the gate-dependence of NHEs in MoSSe evolves under uniaxial strain $u_{xx}$ ranging from $0 \%$ up to $4 \%$(Fig.\ref{FIG03}a). Importantly, we show that a tiny amount of uniaxial strain $u_{xx} \sim 0.2\%$ is sufficient for creating sizeable Berry curvature dipoles of the order $\sim 1 {\AA}$(Fig.\ref{FIG03}b). 

As we explained in subsection A, the uniaxial strain plays the essential role of breaking the $C_3$-symmetry and gives rise to nonzero Berry dipoles in MoSSe. In particular, large Berry dipoles in MoSSe are physically established by the strain-induced energy separation of Berry curvature hot spots associated with left-movers and right-movers(Fig.\ref{FIG01}a). Thus, as one gradually turns on the uniaxial strain, the Berry curvature hot spots associated with left-movers and right-movers start to get separated in energy and nonzero Berry dipole $D_x$ emerges. 

A complete 2D map of $D_x$ as a function of Fermi energy $E_F$ and uniaxial strain $u_{xx}$ is shown in Fig.\ref{FIG03}a, with the black solid line indicating locations of the conduction band minimum(CBM). At $u_{xx} = 0\%$, the $C_3$-symmetry in MoSSe is respected, with $D_x =0$ in all ranges of $E_F$. As $u_{xx}$ is turned on, nonzero Berry dipoles $D_x$ start to emerge. Notably, the band anti-crossing points providing Berry curvature hot spots(Fig.\ref{FIG01}a) are generically located $\sim 20$ meVs away from the CBM. As $E_F$ accesses these hot spots, it is evident from Fig.\ref{FIG03}a that pronounced signals of $D_x$ of the order $\sim 1 {\AA}$ readily appear under very weak uniaxial strains $u_{xx}\sim 0.2\%$. 

To understand why such strong Berry dipoles can be induced by weak strains, we study the evolutionary behavior of $D_x$ under strains by plotting $D_x$ as a function of $E_F$ at various fixed strains: $u_{xx}=0.2\%, 1.2\%, 2.2\%, 3.2\%$, as shown in Fig.\ref{FIG03}b-e. The black double arrows in each case measure the energy separation between the lowest- and highest-lying Berry curvature hot spots. 

As discussed in subsection B, under a uniaxial strain as strong as $u_{xx} = 4\%$, band anti-crossings associated with the left-movers and right-movers are unambiguously separated in energy. As the Berry curvature hot spots are accessed successively, the sign change in the Berry curvature hot spots are signified by the peak-to-dip/dip-to-peak behaviors in the $D_x$-$E_F$ curve(Fig.\ref{FIG02}c). 

Notably, the two dips in $D_x$ originate from the left-movers in the upper band and the right-movers in the lower band respectively, with both their band velocities and Berry curvatures being opposite to each other. As a result, their contributions to $D_x$ are additive(Eq.\ref{eq:BCDipole}). Upon decreasing the uniaxial strain, the two dips start to merge with each other, which enhances the total Berry curvature dipole and reaches a maximum $D_x \sim 2 {\AA}$ for $u_{xx} \sim 1\%$ as shown in Fig.\ref{FIG03}a. By further decreasing $u_{xx}$, while the total $D_x$ would ultimately vanish in the zero strain limit, the additive contributions from merging the two dips in $D_x$ remain strong for weak finite strains. In particular, the magnitude of total $D_x$ is still of the order $\sim 1 {\AA}$ under $u_{xx} \sim 0.2\%$ where the merging between two dips happens(Fig.\ref{FIG03}b). 

It is also worth noting that the special evolutionary behavior of $D_x-E_F$ curves under strains, particularly the merging effect between the two dips upon decreasing the uniaxial strain, provides yet another unique signature for the gate-tunable nonlinear Hall effect in strained MoSSe.

\subsection{\bf{Potential gate-tunable high-frequency rectifiers based on strained polar TMDs}}

In this section, we discuss how the highly tunable NHE in strained polar TMDs provides a potential scheme for switchable high-frequency rectifiers. 

In the past decade, rapid developments of wireless technologies have surged an increasing demand for portable micro-sized devices that can harvest the energy of ambient electromagnetic(EM) radiations. At the heart of these energy harvestors lies the physical process known as rectification, the conversion of oscillating EM fields into DC currents. 

While conventional rectifiers based on semiconductor diodes have found a wide range of industrial applications, a fundamental limitation exists for their operating frequencies\cite{Hemour, Selvan}. In particular, for a prefered current direction to be effectively selected, the diode transition time(\textit{i.e.}, the time scale for a $p-n$ junction to enter a complete open circuit state upon reversing the bias) must be much shorter than the period of EM waves. The typical diode transition time on nanosecond scale sets the maximum frequency limit to lie within the gigahertz range, and the vast amount of energy stored in terahertz and far-infrared radiations, which have natural sources such as thermal radiations, can hardly be harvested with existing rectifiers.

To bypass the frequency threshold, an alternative scheme based on the intrinsic nonlinear property of homogeneous materials was proposed recently\cite{Isobe}. Notably, regardless of the frequency of the applied or ambient AC fields, second-order nonlinearity generically results in a DC response. Thus, the nonlinear Hall effect provides a possible means to build high-frequency rectifiers that can harvest energy of radiations in terahertz and far-infrared regime. In particular, the strained polar TMDs studied in this work can serve as a potential electrically switchable high-frequency rectifier, which controls both the amplitude and direction of rectified currents simply by electric gates. 

With the same set-up in Fig.\ref{FIG02}(d), apart from the second-harmonic component discussed previously, the nonlinear Hall current $\bm{J}_{NH}$ generated by an AC electric field $\bm{E}_{\omega}(t) = \text{Re} \{ \mathcal{E}_x e^{i\omega t} \} \hat{x}$ is partially rectified due to the second-order nonlinearity in $\mathcal{E}_{x}$, with the DC current component given by\cite{Sodemann}:
\begin{eqnarray}\label{eq:CurrentResponse}
j^{0}_y = -\frac{e^3 \tau |\mathcal{E}_x|^2 }{2 (1+i\omega \tau)} D_x.
\end{eqnarray}
It is clear from Eq.\ref{eq:CurrentResponse} that via a moderate gate voltage, the amplitude and direction of the rectified current  $j_{y}^0$ can be regulated in a similar manner as the $D_x-E_F$ curve in Fig.\ref{FIG02}(c). This provides an easy way for charge regulation, which is indispensible for real electronic devices. Thus, strained polar TMDs can be used to build a NHE-based high-frequency rectifier which integrates the generation and regulation of charging DC currents within a single device. 

\section{\bf{Discussion}}

Here, we discuss several important points on nonlinear Hall effects(NHEs) in strained polar TMDs. 

First of all, we point out that while strong spin-orbit interactions are known to exist in TMD materials for years\cite{Xiao, Noah}, the effect of SOCs has been completely ignored in all previous studies on nonlinear Hall physics in strained 2H-TMDs\cite{You, Sodemann}. In this work, we point out for the first time that SOCs can significantly change the nonlinear Hall physics in 2H-TMDs. 

Particularly, in usual strained 2H-TMDs where $\Omega_{spin}$ is absent, the optimal value of $D_x$ due to $\Omega_{orb}$ can only be of order $\sim 0.01 {\AA}$ under a rather strong uniaxial strain $u_{xx} = 2\%$\cite{You, Sodemann}. In sharp contrast, due to $\Omega_{spin}$ induced by SOCs, strong Berry dipole of order $\sim 1 {\AA}$ emerges under a weak uniaxial strain $\sim 0.2\%$(Fig.\ref{FIG03}). 

Moreover, the sign of $D_x$ for electron(hole)-doped samples cannot change due to the fact that the sign of $\Omega_{orb}$ is fixed in both conduction and valence bands\cite{Sodemann, You}. On the other hand, as we show explicitly in Fig.\ref{FIG03}, the nonlinear Hall current becomes highly tunable by gating and strain due to the special property of $\Omega_{spin}$ induced by SOCs. Thus, our work demonstrates that SOCs change the nonlinear Hall physics in TMDs in a qualitative way. Also, as the Berry dipole due to $\Omega_{orb}$ never changes sign, the sign-changing nonlinear Hall current in strained polar TMDs provides a distinctive electrical signature for the recently discovered Berry curvature $\Omega_{spin}$ derived from spin degrees of freedom. 

It is important to note that that NHEs in strained polar TMDs generally have a much stronger gate-dependence than 1T$_d$-WTe$_2$. Strong gate-dependence of NHEs not only provides a practical way to realize gate-tunable Hall devices, but also serves as a distinctive signature of the nontrivial Berry phase origin. In particular, a gate-sensitive nonlinear Hall signal due to Berry curvature dipoles can distinguish itself from trivial mechanisms, such as ratchet effects\cite{Sodemann, Fai, Ganichev}, that are much less sensitive to gating. 

In the recent experiment on bilayer 1T'-WTe$_2$, the sign of the Berry curvature dipole $D_x$ is generally fixed in the neighborhood of the charge neutrality point near the band edges\cite{Pablo}. To switch the sign of $D_x$, one generally needs to gate the Fermi level to at least $50-100$ meVs away from the band edges, which requires a rather strong gating field. Furthermore, without a dual-gate set-up, such a strong gating field inevitably introduces out-of-plane displacement fields that cause complications in band structures as well as Berry curvature effects\cite{Pablo}. To unambiguously identify the Berry phase origin of nonlinear Hall effects in 1T$_d$-WTe$_2$, a dual-gate set-up is necessary to control the carrier density and the displacement field independently.

In contrast, the Berry curvature dipole in strained polar TMDs can switch its sign dramatically within a narrow Fermi level range $\Delta E_F \sim 10-20$ meVs. This range can be easily achieved by a moderate gating, which has been accessed in previous gating experiments on normal 2H-TMDs\cite{Fai2}. Moreover, the strong gate-dependence of $D_x$ in strained polar TMDs occur for Fermi level $\sim 20$ meV measured from the conduction band edge. This corresponds to a relatively low carrier density regime ($n_{2D} \sim 1 \times 10^{12} cm^{-2}$), which has also been readily accessed by weak gating\cite{Fai2, Zefei} without introducing significant displacement field. Thus, a dual-gate set-up is not necessary for detecting the gate-dependence of $D_x$ in strained polar TMDs, and we expect the gate-tunable NHEs in strained polar TMDs to be much more easily observed experimentally comparing to 1T$_d$-WTe$_2$.

Besides, the large magnitude and strong gate-tunability of Berry curvature dipole predicted in this work originates from the intrinsic band anti-crossings caused by Ising and Rashba SOCs in moly-based polar TMDs. It does not require any further experimental design apart from strains. To demonstrate the generality of our prediction, we present the gate-dependence of nonlinear Hall signals in another moly-based polar TMD MoSeTe in Appendix E. Same qualitative features in Fig.\ref{FIG03} are found in the $D_x-E_F$ curve in strained MoSeTe. 

Also, as long as the $C_3$-symmetry is broken, details of the strain configuration do not affect our prediction qualitatively. We further point out that for any strain configuration satisfying $u_{xy} = 0$ and $u_{xx} \neq u_{yy}$, the in-plane mirror symmetry $x \mapsto -x$ is preserved in strained MoSSe, with its point group being identical to bilayer/multilayer 1T$_d$-WTe$_2$\cite{Car, Pablo}. Thus, the Berry curvature dipole has only a nonzero $x$-component $D_x$, which is perpendicular to the mirror plane. In this case, nonlinear Hall effects can be observed as long as the applied electric field deviates from the mirror-invariant $y$-axis, as demonstrated in the recent experiment on multilayer 1T$_d$-WTe$_2$ by Kang et al.\cite{Fai}. 

In addition, it is shown previously that $\Omega_{spin}$ also arises in tungsten(W)-based TMDs\cite{Benjamin}. However, due to the absence of band anti-crossings driven by SOCs in W-based materials, the Berry curvature has a much less non-uniform profile as compared to moly-based case, and the nonlinear Hall effect in W-based polar TMDs is much weaker. Detailed discussions on W-based polar TMDs are presented in Appendix E. 

We note parenthetically that, while a few recent works pointed out possible extrinsic contributions to nonlinear Hall effects due to disorder scattering\cite{Konig, Sodemann2, Du, Cong}, it was suggested that the total nonlinear Hall conductivity remains proportional to the Berry curvature dipole in general\cite{Sodemann2, Du}. In particular, the recent experiment on bilayer 1T'-WTe$_2$ unambiguously demonstrated that the sign of the total nonlinear Hall current is almost strictly controlled by the Berry curvature dipole\cite{Pablo}. Thus, we believe the qualitative features of NHEs in polar TMDs predicted in this work will not be affected by extrinsic effects.

Last but not least, for $p$-type polar TMDs, the effect of $\Omega_{spin}$ is almost negligible near the $K$-points due to the strong Ising SOC $\sim 100$ meVs in the valence band\cite{Liu}, and the total Berry curvature is dominated by the conventional $\Omega_{orb}$\cite{Benjamin}. Thus, for hole-doped polar TMDs, one expects the Berry curvature dipole to be very weak, similar to the case in conventional 2H-TMDs as studied in previous works\cite{Sodemann, You}.

\section*{\bf{Acknowledgement}}

The authors thank Kaifei Kang and K. F. Mak for illuminating discussions. KTL acknowledges the support of Croucher Foundation, Dr. Tai-chin Lo Foundation and HKRGC through C6026-16W, 16309718, 16307117 and 16324216. 

\renewcommand{\theequation}{A-\arabic{equation}}
\renewcommand\thefigure{A-\arabic{figure}}  
\setcounter{equation}{0}  
\setcounter{figure}{0}
\section*{\textbf{Appendix A: Effective Hamiltonian for strained MoSSe}}

\begin{figure}
\centering
\includegraphics[width=3.5in]{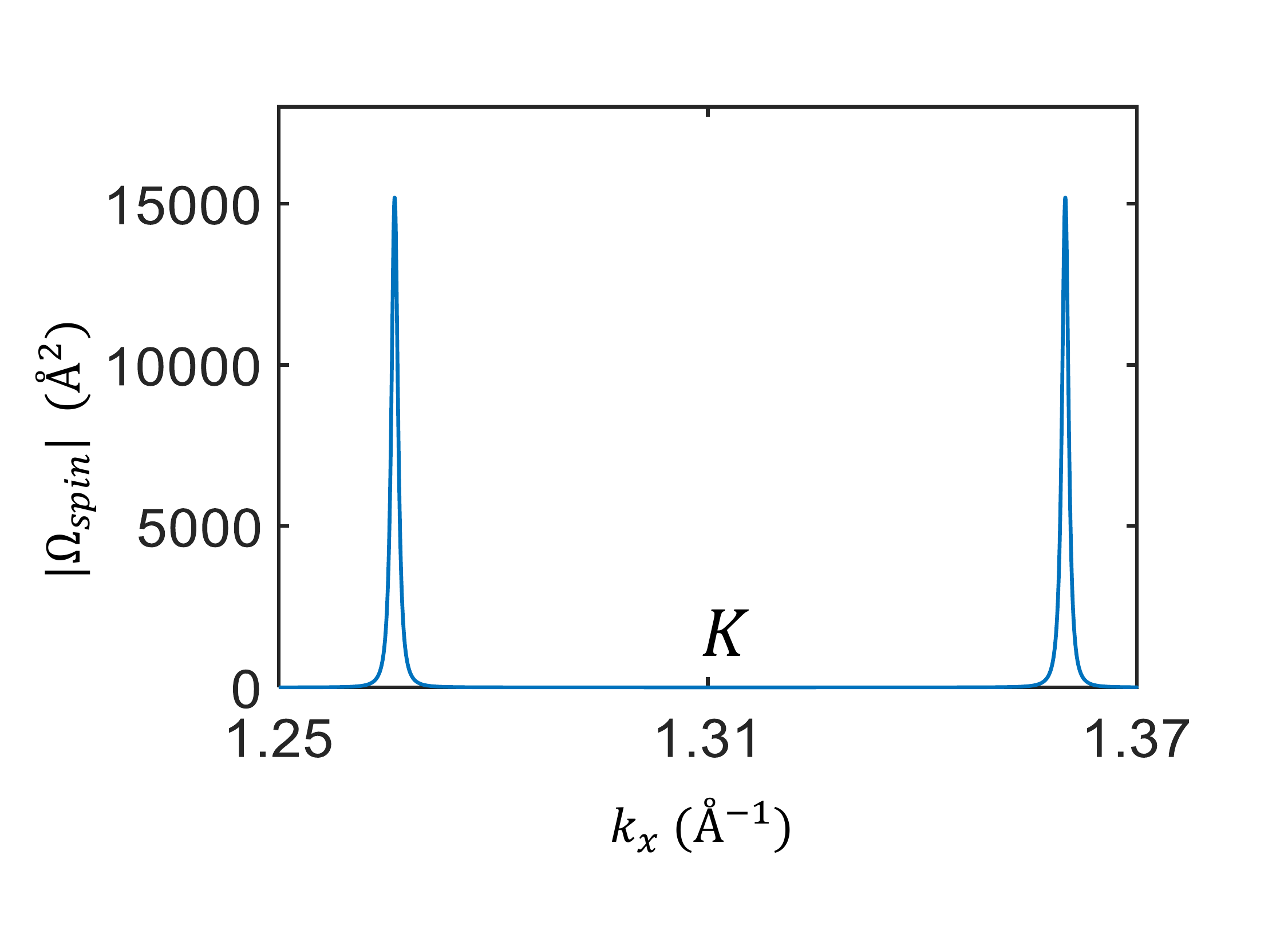}
\caption{Magnitude of Berry curvature generated by Ising and Rashba spin-orbit couplings near $K$-point of unstrained MoSSe. The coordinate of $K$-point is given by $\bm{K} = (4\pi/3a, 0)\approx (1.313, 0)$ with $a = 3.19 \textrm{\AA}$. Given the effective parameters in Table \ref{table:01}, the values of $\Omega_{orb}$ are calculated based on Eq.1-2 of the main text. Evidently, huge $\Omega_{spin}$ arise in the neighborhood of band-anticrossing points $k_0$, with its magnitude generally far greater than $100 \textrm{\AA}^2$.}
\label{FIGS1}
\end{figure}

Here, we derive the effective Hamiltonian near the conduction band minimum for MoSSe by group theory. In the absence of strain, the electronic bands near the conduction band minimum at $K$-points are dominated by the $d_{z^2}$-orbitals, with spin degrees of freedom. As we mentioned in the main text, the symmetry group of MoSSe is given by the product group: $C_{3v} \otimes \mathcal{T}$, with the following generators: (i) three-fold rotation($\hat{C}_{3}$), (ii) mirror reflection($\hat{M}_{x}$) and time-reversal $\mathcal{T}$. The valley index, momentum and spin transform under these generators as follows: 
\begin{eqnarray}\label{eq:01}
\hat{C}_{3} & : & \epsilon \mapsto \epsilon, k_{\pm} \mapsto e^{\pm i2\pi/3}k_{\pm}, \\\nonumber
&  & \sigma_{\pm} \mapsto e^{\pm i2\pi/3}\sigma_{\pm}, \sigma_{z} \mapsto \sigma_{z}.\\\nonumber 
\hat{M}_x & : & \epsilon \mapsto -\epsilon, k_{+}\leftrightarrow-k_{-}, \sigma_{+} \mapsto \sigma_{-}, \sigma_{z} \mapsto -\sigma_{z}. \\\nonumber
\mathcal{T} & : & \epsilon \mapsto -\epsilon, k_{+} \mapsto -k_{-}, \sigma_{+} \mapsto -\sigma_{-}, \sigma_{z} \mapsto -\sigma_{z}.
\end{eqnarray}
where $\epsilon=\pm$ is the valley index, $k_{\pm}=k_{x}\pm ik_{y}$, $\sigma_{\pm}=\sigma_{x}\pm i\sigma_{y}$. The unstrained Hamiltonian $H_0(\bm{k})$ that are invariant under the transformations in (\ref{eq:01}) has the form (up to second order in $\bm{k}$):
\begin{equation}
H_{0}(\bm{k}+\epsilon\bm{K})=\xi_{\bm{k}}\sigma_{0}+\epsilon\beta_{so}(\bm{k})\sigma_{z}+\alpha_{so}(k_{y}\sigma_{x}-k_{x}\sigma_{y}).
\end{equation}
Here $\xi_{\bm{k}}=\frac{k^{2}}{2m^{*}}-\mu$ is the usual kinetic term, $\beta_{so}(\bm{k})=\beta_{0}+\beta_{1}k^{2}$ is the Ising SOC term with $\beta_{0} \beta_{1} < 0$, $\alpha_{so}$ is the strength of Rashba SOC. As we discussed in the main text, in the absence of Rashba SOCs, the spin subbands cross each other at $k_0 = \pm\sqrt{\beta_0 / \beta_1}$. By turning on the Rashba SOC, nontrivial Berry curvatures arise near the $K$-point as shown in Fig.\ref{FIGS1}, with their magnitude being huge in the neighborhood of $k_0$.

Next, we consider corrections terms due to strain. According to Refs.\cite{Shiang}, general strain effects in two-dimensional crystalline solids are mathematically described by a displacement field $\bm{u}(x,y)$ due to distortions in atomic sites in the strained lattice. In particular, physical effects of strains are described by the gradients of $\bm{u}(x,y)$: $\partial_{i} u_{j}, (i,j = x,y)$, which form a second-rank tensor. The four independent components of $\partial_{i} u_{j}$ can be classified according to the irreducible representations of $C_{3v}$: (i) the trace scalar $\partial_{x} u_x + \partial_{y} u_y$, which forms the trivial ($A_1$) representation of $C_{3v}$; (ii) the curl (rotation) of $\bm{u}$: $\overleftrightarrow{\bm{\omega}}_{ij} = \partial_{i} u_j - \partial_{j} u_i$ that forms the $A_{2}$-representation of $C_{3v}$. It describes a rotation of the system about the principal $z$-axis under strain, which can always be taken away by redefining the coordinates; (iii) the symmetric traceless tensor: $\overleftrightarrow{\bm{\epsilon}}_{ij} = \frac{1}{2}[\partial_{i} u_j + \partial_{j} u_i - (\bm{\nabla}\cdot\bm{u})\delta_{ij} ]$, which is characterized by a doublet $\{ \partial_{x} u_y + \partial_{y} u_x, u_{xx} - u_{yy}\}$ that forms the two-dimensional $E$-representation of $C_{3v}$. 

To get rid of the redundant rotational part $\overleftrightarrow{\bm{\omega}}$, the symmetric strain-field tensor $\overleftrightarrow{\bm{u}}_{ij} = \frac{1}{2}(\partial_{i}u_{j} + \partial_{j}u_i)$ is introduced such that strain effects are essentially captured by the independent components of $\overleftrightarrow{\bm{u}}$. Explicitly, under a group element $\hat{g}$, the trace $\partial_{x} u_x + \partial_{y} u_y \equiv u_{xx}+u_{yy}$ remains invariant, while the doublet $\{ \partial_{x} u_y + \partial_{y} u_x, u_{xx} - u_{yy}\} \equiv \{ 2u_{xy}, u_{xx} - u_{yy}\} \equiv \{ u_1, u_2\}$ transforms as a polar vector:
\begin{eqnarray}
\hat{g}:  \{ u_1, u_2\} \mapsto \{ u_1, u_2\} [ D^{(E)}(\hat{g}) ]^{T}.
\end{eqnarray}
where $D^{(E)}(\hat{g})$ denotes the representation matrix for $\hat{g}$ in spatial coordinates: $D^{(E)}_{ij}(\hat{g}) = \bm{e}_{i} \cdot \hat{g} \bm{e}_{j}$, ($i,j=x,y$). Particularly, for the generators $\hat{C}_{3z}, \hat{M}_x$, the representation matrices are:
\begin{eqnarray}
D^{(E)}(\hat{C}_{3z}) &=& 
\begin{pmatrix}
-\frac{1}{2} & -\frac{\sqrt{3}}{2} \\
\frac{\sqrt{3}}{2} & \frac{1}{2}
\end{pmatrix}
,\\\nonumber
D^{(E)}(\hat{M}_{x}) &=& 
\begin{pmatrix}
-1 & 0 \\
0 & 1
\end{pmatrix}.
\end{eqnarray}
To sum up the symmetry properties of the physical quantities mentioned above, the irreducible representations of the valley index, momentum, spin and strain-tensor components are presented in Table \ref{table:00}.

In the presence of the strain-field tensor $\overleftrightarrow{\bm{u}}$, the underlying point group symmetries in $C_{3v}$ is manifested by the following equivalence: Given any $\hat{g} \in C_{3v}$, after being transformed by $\hat{g}$, the system with strain $\overleftrightarrow{\bm{u}}$ is identical to the system with transformed strain $\overleftrightarrow{\bm{u}}' = \hat{g}\overleftrightarrow{\bm{u}}$. Here, components of $\overleftrightarrow{\bm{u}}' $ and $\overleftrightarrow{\bm{u}} $ are related by: $u_0' = u_0, \{ u_1', u_2'\} = \{ u_1, u_2\} [ D^{(E)}(\hat{g}) ]^{T}$. 

Formally, this can be expressed as:
\begin{eqnarray}\label{eq:SymmetryConstraint}
\hat{g}\hat{H}(\overleftrightarrow{\bm{u}})\hat{g}^{-1} = \hat{H}(\hat{g}\overleftrightarrow{\bm{u}}).
\end{eqnarray}

where $\hat{H}(\overleftrightarrow{\bm{u}})$ is the total Hamiltonian under strain $\overleftrightarrow{\bm{u}}$. Note that the symmetry relation above is simply a generalization of the symmetry property of unstrained system with $\overleftrightarrow{\bm{u}} = \bm{0}$. Particularly, Eq.\ref{eq:SymmetryConstraint} implies that the momentum-space Hamiltonian $H(\bm{k})$ satisfies:
\begin{eqnarray}
U^{\dagger}(\hat{g}) H(\hat{g} \bm{k}, \hat{g}\overleftrightarrow{\bm{u}}) U(\hat{g}) = H(\bm{k}, \overleftrightarrow{\bm{u}}) .
\end{eqnarray}

where $U(\hat{g})$ denotes the matrix for group operator $\hat{g}$ under the spin basis: $U_{\sigma \sigma'}(\hat{g}) = \braket{\sigma| \hat{g} | \sigma'}$, with $\sigma, \sigma' = \uparrow, \downarrow$. Accordingly, up to first order in $\bm{k}$, strain effects on the spin-independent terms are described by:
\begin{eqnarray}
H_{strain}(\bm{k}+\epsilon\bm{K}) = [\gamma u_{0}+\epsilon\delta(k_{y}u_{1}-k_{x}u_{2})]\sigma_{0}.
\end{eqnarray}
The $\gamma u_{0}\sigma_{0}$-term describes the on-site energy correction due to strain, which causes an energy offset in the conduction band minimum. The $\epsilon\delta(k_{x}u_{2}-k_{y}u_{1})$-term results from the correction in the bonding strengths due to modified inter-atomic distances in the strained lattice. This term explicitly breaks the $C_{3}$-symmetry, which is responsible for the nonzero Berry curvature dipole discussed in the main text. The total Hamiltonian is 
\begin{equation}
H_{eff}=H_{0}+H_{strain}.
\end{equation}

\begin{table}[ht]
\caption{Irreducible representations(IR)s of valley index, momentum, spin, and strain-tensor components in $C_{3v}$ and parity under time-reversal $\mathcal{T}$.}
\centering
\begin{tabular}{c c c}
\hline \hline
Basis functions & \hspace{0.4 in} IR & \hspace{0.4 in} $\mathcal{T}$ \tabularnewline
\hline 
$\epsilon$ & \hspace{0.4 in} $A_2$ & \hspace{0.4 in} $-$ \tabularnewline
\hline
$1$, $k^2_x + k^2_y$ & \hspace{0.4 in} $A_1$ & \hspace{0.4 in} $+$ \tabularnewline
\hline 
$\{k_x, k_y\}$ & \hspace{0.4 in} $E$ & \hspace{0.4 in} $-$ \tabularnewline
\hline 
$\sigma_0$ & \hspace{0.4 in} $A_1$ & \hspace{0.4 in} $+$ \tabularnewline
\hline
$\sigma_z$ & \hspace{0.4 in} $A_2$ & \hspace{0.4 in} $-$ \tabularnewline
\hline
$\{\sigma_x, \sigma_y\}$ & \hspace{0.4 in} $E$ & \hspace{0.4 in} $-$ \tabularnewline
\hline
$u_0$ & \hspace{0.4 in} $A_1$ & \hspace{0.4 in} $+$ \tabularnewline
\hline
$\{u_1, u_2\}$ & \hspace{0.4 in} $E$ & \hspace{0.4 in} $+$ \tabularnewline
\hline\hline
\end{tabular}
\label{table:00}
\end{table}

The parameters of the effective Hamiltonian can be obtained by fitting the realistic band structures, which are listed in Table \ref{table:01}. Notably, due to $\delta <0$, the uniaxial strain with $u_{xx} > 0, u_{yy} = u_{xy} = 0$ causes a shift of the band minimum at $+K(-K)$ to the left(right), as being consistent with our discussions in the main text. Also, due to $\gamma <0$, the energy offset due to strain is negative for $u_{xx}>0$. By increasing $u_{xx}$, the conduction band minimum gets lower in energy, which is also consistent with the results in Fig.4 of the main text. 

\begin{table}[ht]
\caption{Parameters for the effective Hamiltonian $H_{eff}$.}
\centering
\begin{tabular}{c c c c}
\hline\hline 
$m^{*}/m_{e}$ & \hspace{0.2 in} $\alpha_{so}/a\textrm{(meV)}$ & \hspace{0.2 in} $\beta_{0}\textrm{(meV)}$ & \hspace{0.2 in} $\beta_{1}/a^{2}\textrm{(meV)}$ 
\tabularnewline
\hline 
0.5 & \hspace{0.2 in} 0.3 & \hspace{0.2 in} -1.5 & \hspace{0.2 in} 67 
\tabularnewline
\hline 
$\gamma\textrm{(eV)}$ & \hspace{0.2 in} $\delta/a\textrm{(eV)}$ & \hspace{0.2 in} $a\textrm{(\AA)}$
\tabularnewline
\hline 
-4.2 & \hspace{0.2 in} -1.5 & \hspace{0.2 in} 3.19
\tabularnewline
\hline\hline 
\end{tabular}
\label{table:01}
\end{table}

\renewcommand{\theequation}{B-\arabic{equation}}
\setcounter{equation}{0}  
\section*{\textbf{Appendix B: Tight-binding Hamiltonian}}\label{AppendixB}

The tight-binding(TB) Hamiltonian for MoS$_{2}$ and MoSSe takes the same form, with the only difference that the Rashba spin-orbit coupling(SOC) for pristine MoS$_{2}$ is zero. In generic monolayer transition-metal dichalcogenides, the conduction and valence band edges are dominated by the $d_{z^2}$, $d_{xy}$, $d_{x^2-y^2}$ orbitals from the transition-metal atoms\cite{Liu}. In the spinful Bloch basis of the $d$-orbitals: $\{ \ket{\bm{k},d_{z^2}}, \ket{{\bm{k}, d_{xy}}}, \ket{\bm{k}, d_{x^2-y^2}}\}$, the unstrained tight-binding Hamiltonian for polar TMD up to nearest-neighbor hopping is written as
\begin{eqnarray}
H^{0}_{\text{TB}}\left(\bm{k}\right)&=&H_{\text{NN}}\left(\bm{k}\right)\otimes\sigma_0+\frac{1}{2}\lambda L_z\otimes\sigma_z \\\nonumber
&+& H_{R}(\bm{k}) + H^{c}_{I}(\bm{k}).
\end{eqnarray}
The first term $H_{\text{NN}}\left(\bm{k}\right)$ represents the spin-independent terms, the second term refers to the atomic spin-orbit coupling. $H_{R}(\bm{k})$ and $H^{c}_{I}(\bm{k})$ describe the Rashba SOCs and the Ising SOC near the conduction band edges. The first two terms are given by\cite{Liu}:
\begin{eqnarray}
H_{\text{NN}}\left(\bm{k}\right)&=&
\begin{pmatrix}
V_0 & V_1 & V_2 \\ 
V_1^{\ast} & V_{11} & V_{12} \\ 
V_2^{\ast} & V_{12}^{\ast} & V_{22}
\end{pmatrix} - \mu I_{3\times3}, \\\nonumber
 L_z&=&
 \begin{pmatrix}
0 & 0 & 0 \\ 
0 & 0 & -2i \\ 
0 & 2i & 0
\end{pmatrix}
\end{eqnarray}
Here, $\mu$ denotes the chemical potential, and $L_z$ is the $z$-component of the orbital angular momentum. Defining $\left(\alpha, \beta\right)=\left(\frac{1}{2}k_xa, \frac{\sqrt{3}}{2}k_ya\right)$, $V_0$, $V_1$, $V_2$, $V_{11}$, $V_{12}$ and $V_{22}$ are expressed as:
\begin{eqnarray}
V_0 & = & \epsilon_1+2t_0\left(2\cos\alpha\cos\beta+\cos 2\alpha\right),
\end{eqnarray}
\begin{eqnarray}
\text{Re}\left[V_1\right] & = & -2\sqrt{3}t_2\sin\alpha\sin\beta,
\end{eqnarray}
\begin{eqnarray}
\text{Im}\left[V_1\right] & = & 2t_1\sin\alpha\left(2\cos\alpha+\cos\beta\right),
\end{eqnarray}
\begin{eqnarray}
\text{Re}\left[V_2\right] & = & 2t_2\left(\cos2\alpha-\cos\alpha\cos\beta\right),
\end{eqnarray}
\begin{eqnarray}
\text{Im}\left[V_2\right] & = & 2\sqrt{3}t_1\cos\alpha\sin\beta,
\end{eqnarray}
\begin{eqnarray}
V_{11}&=&\epsilon_2+\left(t_{11}+3t_{22}\right)\cos\alpha\cos\beta \\\nonumber
&+& 2t_{11}\cos2\alpha,
\end{eqnarray}
\begin{eqnarray}
\text{Re}\left[V_{12}\right]&=&\sqrt{3}\left(t_{22}-t_{11}\right)\sin\alpha\sin\beta,
\end{eqnarray}
\begin{eqnarray}
\text{Im}\left[V_{12}\right]&=&4t_{12}\sin\alpha\left(\cos\alpha-\cos\beta\right),
\end{eqnarray}
and
\begin{eqnarray}
V_{22}&=&\epsilon_2+\left(3t_{11}+t_{22}\right)\cos\alpha\cos\beta \\\nonumber
&+& 2t_{22}\cos2\alpha.
\end{eqnarray}

The parameters for the NN tight-binding model are adapted from Refs.\cite{Liu} and listed in Table \ref{table:02}.

\begin{widetext}
\begin{center}
\begin{table}[ht]
\caption{Parameters for $H_{\text{NN}}\left(\bm{k}\right)$ for monolayer MoS$_2$ and MoSe$_2$ adapted from Refs.\cite{Liu}. All energy parameters set in units of $eV$.} % title of Table
\centering % used for centering table
\begin{tabular}{c c c c c c c c c c c c c} % centered columns
\hline\hline %inserts double horizontal lines
&& $a(\AA)$ & $\epsilon_1$ & $\epsilon_2$ & $t_0$ & $t_1$ & $t_2$ & $t_{11}$ & $t_{12}$ & $t_{22}$ & $\lambda$\\ [0.5ex] % inserts table
%heading
\hline % inserts single horizontal line
MoS$_2$ && 3.190 & 1.046 & 2.104 & -0.184 & 0.401 & 0.507 & 0.218 & 0.338 & 0.057 & -0.073 \\% inserting body of the table
[0.5ex] % [1ex] adds vertical space
MoSe$_2$ && 3.326 & 0.919 & 2.065 & -0.188 & 0.317 & 0.456 & 0.211 & 0.290 & 0.130 & -0.091 \\ 
\hline\hline %inserts double line
\end{tabular}
\label{table:02} % is used to refer this table in the text
\end{table}
\end{center}
\end{widetext}

Next, we present the Ising and Rashba SOCs in the tight-binding model. Note that the Ising SOCs in the valence bands are readily described by $H_{\text{NN}}(\bm{k})$ together with the atomic spin-orbit coupling term. The Ising SOC in the conduction bands $H^{c}_{I}(\bm{k})$ takes the form:
\begin{eqnarray}
H^{c}_{I}(\bm{k})=
\begin{pmatrix}
\beta(\bm{k}) & 0 & 0 \\ 
0 & 0 & 0 \\ 
0 & 0 & 0
\end{pmatrix}
\otimes\sigma_z.
\end{eqnarray}
where
\begin{eqnarray}
\beta({\bm{k}}) = -\frac{2\beta^{c}_{so}}{3\sqrt{3}}[\sin(2\alpha)-2\sin(\alpha)\cos(\beta)].
\end{eqnarray}
with $\beta^{c}_{so}=-1.5$ meV. The Rashba SOC $H_{R}(\bm{k})$ is written as:
\begin{eqnarray}
H_{R}(\bm{k})=
\begin{pmatrix}
2\alpha_0 & 0 & 0 \\ 
0 & 0 & 0 \\ 
0 & 0 & 0
\end{pmatrix}
\otimes(f_x(\bm{k})\sigma_y - f_y(\bm{k})\sigma_x).
\end{eqnarray}
where $\alpha_0$ denotes the tight-binding Rashba parameter for $d_{z^2}$-orbitals. In our tight-binding calculations, we set $\alpha_0=0.2$ meV for MoSSe. We note that $\alpha_0$ is related to the effective Rashba strength $\alpha_{so}$ in the effective model by: $\alpha_{so}= \frac{3}{2} \alpha_0 a$, where $a$ is the lattice constant. The Rashba SOC for $\{d_{xy}, d_{x^2-y^2}\}$ orbitals are neglected as we only concern about the conduction band minimum which is dominated by $d_{z^2}$-orbitals. The functions $f_x(\bm{k}), f_y(\bm{k})$ are given by:
\begin{eqnarray}
f_x(\bm{k}) &=& \sin(2\alpha)+\sin(\alpha)\cos(\beta) \\\nonumber
f_y(\bm{k}) &=& \sqrt{3}\sin(\beta)\cos(\alpha)
\end{eqnarray}

\renewcommand{\theequation}{C-\arabic{equation}}
\setcounter{equation}{0}  
\section*{\textbf{Appendix C: Symmetry-allowed linear coupling to strain-field}}

Considering $\{d_{z^2}, d_{xy}, d_{x^2-y^2}\}$-orbitals, the general form of real-space tight-binding Hamiltonian is given by:
\begin{eqnarray}
\hat{H}(\overleftrightarrow{\bm{u}}) = \sum_{\bm{R}, \bm{R}'} \sum_{\alpha,\beta} c^{\dagger}_{\alpha} (\bm{R}) h_{\alpha,\beta} (\bm{R}, \bm{R}', \overleftrightarrow{\bm{u}}) c^{\dagger}_{\beta} (\bm{R}')
\end{eqnarray}
where $\bm{R}, \bm{R'}$ label the lattice sites, and $\alpha, \beta$ are the orbital(spin) indices. Based on the general relation defined in Eq.\ref{eq:SymmetryConstraint} of Section I, the matrix connecting a general pair of lattice sites $\bm{R}, \bm{R'}$ satisfy the following relation:
\begin{eqnarray}\label{eq:PointGroupRelation}
\hat{h}(\hat{g}\bm{R}, \hat{g}\bm{R}', \overleftrightarrow{\bm{u}}) = U(\hat{g}) \hat{h} (\bm{R},\bm{R}', \hat{g}^{-1}\overleftrightarrow{\bm{u}}) U^{\dagger}(\hat{g}).
\end{eqnarray}

Note that Eq.\ref{eq:PointGroupRelation} is simply a generalization of the symmetry relations for unstrained case as studied in Refs.\cite{Liu}, should we enforce that under $\hat{g}$ the strain-field tensor transforms inversely to $\overleftrightarrow{\bm{u}}' = \hat{g}^{-1}\overleftrightarrow{\bm{u}}$, with $u_0' = u_0, \{ u_1', u_2'\} = \{ u_1, u_2\} [ D^{(E)}(\hat{g}^{-1}) ]^{T}$. 

Under weak applied strains considered in this work, it is reasonable to assume that corrections due to strain are essentially given by linear-order coupling to the strain-field components in $\overleftrightarrow{\bm{u}}$. In particular, the on-site terms are invariant under the generalized transformation in Eq.\ref{eq:PointGroupRelation}, which take the following form:
\begin{eqnarray}\label{eq:InvariantForm}
\hat{h}_{0}(\overleftrightarrow{\bm{u}}) &=& (u_{xx} + u_{yy}) 
\begin{pmatrix}
E_0^{S} & 0 & 0\\
0 &  E_2^{S} & 0\\
0 &  0 & E_2^{S}
\end{pmatrix}\\\nonumber
&+& 2 u_{xy}
\begin{pmatrix}
0 & h^{S}_1 & 0\\
h^{S}_1 & 0& h^{S}_2\\
0 & h^{S}_2 & 0
\end{pmatrix}\\\nonumber
&+& (u_{xx} - u_{yy}) 
\begin{pmatrix}
0 &0 & h^{S}_1\\
0 & h^{S}_2 & 0\\
h^{S}_1 & 0 & -h^{S}_2
\end{pmatrix}.
\end{eqnarray}

Next, let us consider nearest-neighbor terms. According to Eq.\ref{eq:PointGroupRelation}, given the hopping term $\hat{h}(\bm{\delta}, \overleftrightarrow{\bm{u}})$ along a certain bonding vector $\bm{\delta} = \bm{R} - \bm{R}'$, the hopping Hamiltonian along the transformed bonding vector $\bm{\delta}' = \hat{g}\bm{\delta}$ can be explicitly given by:
\begin{eqnarray}
\hat{h} (\bm{\delta}', \overleftrightarrow{\bm{u}}) = U(\hat{g}) \hat{h}(\bm{\delta}, \hat{g}^{-1} \overleftrightarrow{\bm{u}}) U^{\dagger}(\hat{g})
\end{eqnarray}
 
Thus, with the knowledge of the hopping Hamiltonian $\hat{h}(\bm{\delta}, \overleftrightarrow{\bm{u}})$ along $\bm{\delta}$, hopping terms along all bonding vectors $\bm{\delta}'=\hat{g}\bm{\delta}$ can be obtained by acting $\hat{g}$'s on $\hat{h}(\bm{\delta}, \overleftrightarrow{\bm{u}})$. The symmetry allowed form for $\hat{h} (\bm{\delta}_1 \equiv a\hat{x}, \overleftrightarrow{\bm{u}})$ is given by\cite{Shiang}:
\begin{eqnarray}\label{eq:hdeltax}
\hat{h} (\bm{\delta}_1, \overleftrightarrow{\bm{u}}) &=& u_0 \hat{P} ^{(A_1)} + u_1 \hat{N} ^{(E)} + u_2 \hat{P} ^{(E)}\\\nonumber
& = & (u_{xx} + u_{yy})
\begin{pmatrix}
P^{(A_1)}_{00} & P^{(A_1)}_{01} & P^{(A_1)}_{02}\\
-P^{(A_1)}_{01} & P^{(A_1)}_{11} & P^{(A_1)}_{12}\\
P^{(A_1)}_{02} & -P^{(A_1)}_{12} & P^{(A_1)}_{22}\\
\end{pmatrix}\\\nonumber
 &+&  (u_{xx} - u_{yy})
\begin{pmatrix}
P^{(E)}_{00} & P^{(E)}_{01} & P^{(E)}_{02}\\
-P^{(E)}_{01} & P^{(E)}_{11} & P^{(E)}_{12}\\
P^{(E)}_{02} & -P^{(E)}_{12} & P^{(E)}_{22}\\
\end{pmatrix}
\\\nonumber
&+& 2u_{xy}
\begin{pmatrix}
0 & N^{(E)}_{01} & N^{(E)}_{02}\\
N^{(E)}_{01} & 0 & N^{(E)}_{12}\\
-N^{(E)}_{02} & N^{(E)}_{12} & 0\\
\end{pmatrix}.
\end{eqnarray}

The bonds in the direction $\bm{\delta}' = \hat{C}_{3z}\bm{\delta}$ are related to the bonds in the direction $\bm{\delta}$ by\cite{Shiang}:
\begin{eqnarray}
\hat{h} (\bm{\delta}', \overleftrightarrow{\bm{u}}) = U(\hat{C}_{3z}) \hat{h}(\bm{\delta}, \hat{C}_{3z}^{-1} \overleftrightarrow{\bm{u}} ) U^{\dagger}(\hat{C}_{3z})
\end{eqnarray}
where
\begin{eqnarray}
U(\hat{C}_{3z})  =
\begin{pmatrix}
1 & 0 & 0 \\ 0 & -\frac{1}{2} & -\frac{\sqrt{3}}{2} \\ 0 & \frac{\sqrt{3}}{2} & -\frac{1}{2}
\end{pmatrix} 
\end{eqnarray}
And for the transformed strain tensor field $\overleftrightarrow{\bm{u}}' = \hat{C}_{3z}^{-1} \overleftrightarrow{\bm{u}}$, we have:
\begin{eqnarray}
u'_{xx} - u'_{yy} &=& -\frac{1}{2} (u_{xx} - u_{yy}) -\frac{\sqrt{3}}{2} (2u_{xy})\\\nonumber
2 u'_{xy} &=& -\frac{1}{2} (2 u_{xy} ) + \frac{\sqrt{3}}{2} (u_{xx} - u_{yy})
\end{eqnarray}
in accord with the $E$-representation of $C_{3v}$. Now, with the general form of $\hat{h} (\bm{\delta}_1, \overleftrightarrow{\bm{u}})$ in Eq.\ref{eq:hdeltax}, we obtain the hopping terms in  $\bm{\delta}_2 \equiv \hat{C}_{3z}\bm{\delta}_1 $ and $\bm{\delta}_3 \equiv \hat{C}^{2}_{3z}\bm{\delta}_1$ using the relations above, which are given by:

\begin{widetext}
\begin{eqnarray}\label{eq:hdelta2}
\hat{h} (\bm{\delta}_2, \overleftrightarrow{\bm{u}}) &=& (u_{xx} + u_{yy})
\begin{pmatrix}
P^{(A1)}_{00} & -\frac{1}{2} P^{(A1)}_{01} - \frac{\sqrt{3}}{2} P^{(A1)}_{02}  & \frac{\sqrt{3}}{2}P^{(A1)}_{01} - \frac{1}{2} P^{(A1)}_{02}\\
\frac{1}{2} P^{(A1)}_{01} - \frac{\sqrt{3}}{2} P^{(A1)}_{02} & \frac{1}{4}P^{(A1)}_{11} + \frac{3}{4}P^{(A1)}_{22} & P^{(A1)}_{12} + \frac{\sqrt{3}}{4} (P^{(A1)}_{22} - P^{(A1)}_{11})\\
-\frac{\sqrt{3}}{2}P^{(A1)}_{01} - \frac{1}{2} P^{(A1)}_{02} & -P^{(A1)}_{12} + \frac{\sqrt{3}}{4} (P^{(A1)}_{22} - P^{(A1)}_{11}) & \frac{3}{4}P^{(A1)}_{11} + \frac{1}{4}P^{(A1)}_{22}\\
\end{pmatrix}
\\\nonumber
&+& [-\frac{\sqrt{3}}{2} (2u_{xy}) -\frac{1}{2} (u_{xx} - u_{yy})]
\begin{pmatrix}
P^{(E)}_{00} & -\frac{1}{2} P^{(E)}_{01} - \frac{\sqrt{3}}{2} P^{(E)}_{02}  & \frac{\sqrt{3}}{2}P^{(E)}_{01} - \frac{1}{2} P^{(E)}_{02}\\
\frac{1}{2} P^{(E)}_{01} - \frac{\sqrt{3}}{2} P^{(E)}_{02} & \frac{1}{4}P^{(E)}_{11} + \frac{3}{4}P^{(E)}_{22} & P^{(E)}_{12} + \frac{\sqrt{3}}{4} (P^{(E)}_{22} - P^{(E)}_{11})\\
-\frac{\sqrt{3}}{2}P^{(E)}_{01} - \frac{1}{2} P^{(E)}_{02} & -P^{(E)}_{12} + \frac{\sqrt{3}}{4} (P^{(E)}_{22} - P^{(E)}_{11}) & \frac{3}{4}P^{(E)}_{11} + \frac{1}{4}P^{(E)}_{22}\\
\end{pmatrix}
\\\nonumber
&+& [-\frac{1}{2} (2u_{xy}) +\frac{\sqrt{3}}{2} (u_{xx} - u_{yy})]
\begin{pmatrix}
0 & -\frac{1}{2} N^{(E)}_{01} - \frac{\sqrt{3}}{2} N^{(E)}_{02}  & \frac{\sqrt{3}}{2}N^{(E)}_{01} - \frac{1}{2} N^{(E)}_{02}\\
-\frac{1}{2} N^{(E)}_{01} + \frac{\sqrt{3}}{2} N^{(E)}_{02} & \frac{\sqrt{3}}{2}N^{(E)}_{12} & -\frac{1}{2}N^{(E)}_{12}\\
 \frac{\sqrt{3}}{2}N^{(E)}_{01} + \frac{1}{2} N^{(E)}_{02} & -\frac{1}{2}N^{(E)}_{12} & -\frac{\sqrt{3}}{2}N^{(E)}_{12}\\
\end{pmatrix}
\end{eqnarray}

\begin{eqnarray}\label{eq:hdelta3}
\hat{h} (\bm{\delta}_3, \overleftrightarrow{\bm{u}}) &=& (u_{xx} + u_{yy})
\begin{pmatrix}
P^{(A1)}_{00} & -\frac{1}{2} P^{(A1)}_{01} + \frac{\sqrt{3}}{2} P^{(A1)}_{02}  & -\frac{\sqrt{3}}{2}P^{(A1)}_{01} - \frac{1}{2} P^{(A1)}_{02}\\
\frac{1}{2} P^{(A1)}_{01} + \frac{\sqrt{3}}{2} P^{(A1)}_{02} & \frac{1}{4}P^{(A1)}_{11} + \frac{3}{4}P^{(A1)}_{22} & P^{(A1)}_{12} - \frac{\sqrt{3}}{4} (P^{(A1)}_{22} - P^{(A1)}_{11})\\
\frac{\sqrt{3}}{2}P^{(A1)}_{01} - \frac{1}{2} P^{(A1)}_{02} & -P^{(A1)}_{12} - \frac{\sqrt{3}}{4} (P^{(A1)}_{22} - P^{(A1)}_{11}) & \frac{3}{4}P^{(A1)}_{11} + \frac{1}{4}P^{(A1)}_{22}\\
\end{pmatrix}
\\\nonumber
&+& [\frac{\sqrt{3}}{2} (2u_{xy}) -\frac{1}{2} (u_{xx} - u_{yy})]
\begin{pmatrix}
P^{(E)}_{00} & -\frac{1}{2} P^{(E)}_{01} + \frac{\sqrt{3}}{2} P^{(E)}_{02}  & -\frac{\sqrt{3}}{2}P^{(E)}_{01} - \frac{1}{2} P^{(E)}_{02}\\
\frac{1}{2} P^{(E)}_{01} + \frac{\sqrt{3}}{2} P^{(E)}_{02} & \frac{1}{4}P^{(E)}_{11} + \frac{3}{4}P^{(E)}_{22} & P^{(E)}_{12} - \frac{\sqrt{3}}{4} (P^{(E)}_{22} - P^{(E)}_{11})\\
\frac{\sqrt{3}}{2}P^{(E)}_{01} - \frac{1}{2} P^{(E)}_{02} & -P^{(E)}_{12} - \frac{\sqrt{3}}{4} (P^{(E)}_{22} - P^{(E)}_{11}) & \frac{3}{4}P^{(E)}_{11} + \frac{1}{4}P^{(E)}_{22}\\
\end{pmatrix}
\\\nonumber
&+& [-\frac{1}{2} (2u_{xy}) -\frac{\sqrt{3}}{2} (u_{xx} - u_{yy})]
\begin{pmatrix}
0 & -\frac{1}{2} N^{(E)}_{01} + \frac{\sqrt{3}}{2} N^{(E)}_{02}  & -\frac{\sqrt{3}}{2}N^{(E)}_{01} - \frac{1}{2} N^{(E)}_{02}\\
-\frac{1}{2} N^{(E)}_{01} - \frac{\sqrt{3}}{2} N^{(E)}_{02} & -\frac{\sqrt{3}}{2}N^{(E)}_{12} & -\frac{1}{2}N^{(E)}_{12}\\
 -\frac{\sqrt{3}}{2}N^{(E)}_{01} + \frac{1}{2} N^{(E)}_{02} & -\frac{1}{2}N^{(E)}_{12} & \frac{\sqrt{3}}{2}N^{(E)}_{12}\\
\end{pmatrix}
\end{eqnarray}
\end{widetext}

The Hermitian property of $H$ implies that: $h (\bm{\delta}, \overleftrightarrow{\bm{u}}) = h^{\dagger} (-\bm{\delta}, \overleftrightarrow{\bm{u}})$, thus the rest of the hopping terms in $-\bm{\delta}_i  (i=1,2,3)$ can simply be obtained by hermitian conjugation. With hopping terms along all bonding vectors above, we Fourier transform the Wannier operators:
\begin{eqnarray}
\hat{c}^{\dagger}_{\alpha}(\bm{R}) =  \frac{1}{\sqrt{N}} \sum_{\bm{k}} e^{-i \bm{k} \cdot \bm{R} } \hat{c}^{\dagger}_{\bm{k} \alpha}
\end{eqnarray}

and the hopping terms in momentum-space become
\begin{eqnarray}
h_1(\bm{k}, \overleftrightarrow{\bm{u}}) = \sum_{n=1}^{3} e^{-i \bm{k}\cdot \bm{\delta}_{n}} \hat{h} (\bm{\delta}_n, \overleftrightarrow{\bm{u}}) + h.c.
\end{eqnarray}

The strained momentum-space Hamiltonian can be obtained as:
\begin{eqnarray}
H_{\text{TB}}^S (\bm{k}) &=& h_0(\overleftrightarrow{\bm{u}}) \otimes \sigma_0 + h_1(\bm{k}, \overleftrightarrow{\bm{u}})\otimes \sigma_0.
\end{eqnarray}

Thus, the total tight-binding Hamiltonian for strained 2H-TMDs is given by:
\begin{eqnarray}
H_{\text{TB}}(\bm{k}) = H_{\text{TB}}^0 (\bm{k}) + H_{\text{TB}}^S (\bm{k}).
\end{eqnarray}
The model parameters for $H_{\text{TB}}(\bm{k})$ used in calculating Fig.2 $\&$ 3 of the main text are presented in Table \ref{table:02}(unstrained parameters) and Table \ref{table:03}(strained parameters).

\begin{widetext}
\begin{center}
\begin{table}[ht]
\caption{Strained tight-binding parameters up to nearest-neighbor(NN) hopping terms for monolayer MoS$_2$ and MoSe$_2$ adapted from Refs.\cite{Shiang}. All parameters are set in units of eV.} % title of Table
\centering % used for centering table
\begin{tabular}{c c c c c c c c | c c c c c c c c c} % centered columns
\hline\hline %inserts double horizontal lines
&&&& MoS$_2$ &&&&&&&&& MoSe$_2$ \\[0.5ex]
\hline
On-site && $E^{S}_{0}$ & $E^{S}_{2}$ & $h^{S}_{1}$ & $h^{S}_{2}$ 
 &&&&&& $E^{S}_{0}$ & $E^{S}_{2}$ & $h^{S}_{1}$ & $h^{S}_{2}$ \\ [0.5ex] % inserts table
\hline
&& -1.021 &  -1.817 & -0.043 & -0.370
&&&&&& -1.090 &  -2.023 & 0.004 & -0.296\\ [0.5ex]
%heading
\hline % inserts single horizontal line
NN && $P^{(A1)}_{00}$ & $P^{(A1)}_{01}$ & $P^{(A1)}_{02}$ & $P^{(A1)}_{11}$ & $P^{(A1)}_{12}$ & $P^{(A1)}_{22}$ 
 &&&& $P^{(A1)}_{00}$ & $P^{(A1)}_{01}$ & $P^{(A1)}_{02}$ & $P^{(A1)}_{11}$ & $P^{(A1)}_{12}$ & $P^{(A1)}_{22}$\\ [0.5ex] % inserts table
%heading
\hline % inserts single horizontal line
 && 1.032 &  -0.285 & -0.738 & -1.027 & 0.206 & 1.544 
 &&&& 0.885 &  -0.236 & -0.596 & -0.951 & 0.195 & 1.333 \\ [1ex] % [1ex] adds vertical space
\hline
&& $P^{(E)}_{00}$ & $P^{(E)}_{01}$ & $P^{(E)}_{02}$ & $P^{(E)}_{11}$ & $P^{(E)}_{12}$ & $P^{(E)}_{22}$ 
&&&& $P^{(E)}_{00}$ & $P^{(E)}_{01}$ & $P^{(E)}_{02}$ & $P^{(E)}_{11}$ & $P^{(E)}_{12}$ & $P^{(E)}_{22}$ \\ [1ex]
\hline
&& 0.376 & -0.188 & -0.779 & -0.910 & -0.003 & 1.337 
&&&& 0.333 & -0.126 & -0.667 & -0.793 & 0.008 & 1.108 \\ [1ex]
\hline
&& $N^{(E)}_{01}$ & $N^{(E)}_{02}$ & $N^{(E)}_{12}$ 
&&&&&&& $N^{(E)}_{01}$ & $N^{(E)}_{02}$ & $N^{(E)}_{12}$ \\ [1ex]
\hline
&& 0.288 & 0.152 & -0.634 
&&&&&&& 0.255 & 0.110 & -0.565 \\ [1ex]
\hline
\end{tabular}
\label{table:03} % is used to refer this table in the text
\end{table}

\end{center}
\end{widetext}

\renewcommand{\theequation}{D-\arabic{equation}}
\renewcommand\thefigure{D-\arabic{figure}}  
\setcounter{equation}{0}  
\setcounter{figure}{0}
\section*{\textbf{Appendix D: Strongly enhanced Berry dipole due to SOCs}}

In the main text, we mentioned that $\Omega_{spin}$ not only enhances the magnitude of total Berry curvature $\Omega_{tot}$ but also results in a highly nonuniform momentum-space profile for $\Omega_{tot}$. In this section, we discuss in details how $\Omega_{spin}$ leads to $\Omega_{tot}$ with very large magnitudes and nonuniform profile, and then explain why these special properties result in strongly enhanced Berry curvature dipole in strained MoSSe. 

First of all, we show that, in the regime where large Berry dipoles emerge (Fermi level lying close to the band anti-crossing points), the total Berry curvature $\Omega_{tot}$ in the electron bands can be approximately given as an algebraic sum of $\Omega_{spin}$ and $\Omega_{orb}$, \textit{i.e.}, $\Omega_{tot} \approx \Omega_{spin} + \Omega_{orb}$. 

To demonstrate the relation $\Omega_{tot} \approx \Omega_{spin} + \Omega_{orb}$, we consider $K = (4\pi/3a,0)$ with the physics near $-K$ followed by time-reversal symmetry. In the Bloch basis of $\ket{d_{\textrm{c},\uparrow}}, \ket{d_{\textrm{c},\downarrow}}, \ket{d_{\textrm{v},\uparrow}}, \ket{d_{\textrm{v},\downarrow}}$, the four-band Hamiltonian near $K$ can be written as\cite{Xiao, Benjamin}:
\begin{widetext}
\begin{eqnarray}\label{eq:FourbandH}
H_{tot}(\bm{k}) =
\begin{pmatrix}
\frac{\Delta}{2} + (\beta_0 + \beta'_1k^2) & -i\alpha_{so} (k_x +ik_y) & V_F (k_x -ik_y) & 0\\
i\alpha_{so} (k_x -ik_y) & \frac{\Delta}{2} - (\beta_0 + \beta'_1k^2) & 0 & V_F (k_x -ik_y) \\
 V_F (k_x +ik_y) & 0 & -\frac{\Delta}{2} + \lambda & 0\\
 0 & V_F (k_x +ik_y) & 0 & -\frac{\Delta}{2} - \lambda
\end{pmatrix}.
\end{eqnarray}
\end{widetext}
Here, $\ket{d_{\textrm{c}, \alpha}} \equiv \ket{d_{z^2},\alpha}$ denotes the predominant $\ket{d_{z^2}}$-orbital with spin $\alpha = \uparrow, \downarrow$ at the conduction band edge at $K$, and $\ket{d_{\textrm{v}, \alpha}} \equiv \ket{d_{x^2-y^2} + i d_{xy}, \alpha}$ denotes the predominant $\ket{d_{x^2-y^2} + i d_{xy}}$-orbital with spin $\alpha = \uparrow, \downarrow$ at the valence band edge at $K$. $V_F = 3.5 eV\cdot {\AA}$ is the effective inter-orbital hopping parameter, $\Delta = 1.66 eV$ is the band gap at $K$, and $\lambda \approx 50$ meVs describes the Ising SOC strength in the valence band. The parameters $\beta_0, \alpha_{so}$ are defined in the same way as in the main text. However, $\beta'_1$ is not to be identified with $\beta_1$ in Eq.1-2 of the main text. The relation between $\beta_1$ and the parameters $\beta'_1$, $V_F$, $\Delta$, $\lambda$ will become clear in the following discussions.

Without loss of generality, we demonstrate the case of the lower SOC-split conduction band (band index: $n = \textrm{c},-$). The analysis for the upper SOC-split conduction band ($n = \textrm{c},+$) is similar. For any momentum $\bm{k}$ displaced from $K$, the \textit{exact} form of total Berry curvature in the lower conduction band is given by:
\begin{widetext}
\begin{eqnarray}{\label{eq:Omegatot}}
\Omega^{tot}_{\textrm{c},-}(\bm{k}) &=& \Omega^{inter}_{\textrm{c},-}(\bm{k}) + \Omega^{intra}_{\textrm{c},-}(\bm{k}), \\\nonumber
\Omega^{inter}_{\textrm{c},-}(\bm{k}) &=& i \sum_{m=\pm}\frac{\braket{\textrm{c},-,\bm{k}|\hat{v}_x|\textrm{v},m,\bm{k}} \braket{\textrm{v},m,\bm{k}|\hat{v}_y|\textrm{c},-,\bm{k}} - c.c.} {(E_{\textrm{c},-}(\bm{k}) - E_{\textrm{v},m}(\bm{k}))^2}, \\\nonumber
\Omega^{intra}_{\textrm{c},-}(\bm{k}) &=& i \frac{\braket{\textrm{c},-,\bm{k}|\hat{v}_x|\textrm{c},+,\bm{k}} \braket{\textrm{c},+,\bm{k}|\hat{v}_y|\textrm{c},-,\bm{k}} - c.c.} {(E_{\textrm{c},-}(\bm{k}) - E_{\textrm{c},+}(\bm{k}))^2}.
\end{eqnarray}
\end{widetext}
Here, $n = \textrm{c/v}, +/-$ is the band index for Bloch eigenstate $\ket{n,\bm{k}}$ of $H_{tot}(\bm{k})$, associated with band energy $E_n(\bm{k})$. $+/-$ labels the upper/lower SOC-split bands, and $\textrm{c/v}$ labels the conduction/valence band. $\hat{v}_{i} \equiv \partial H_{tot} /\partial k_i$, with $i = x,y$, is the velocity operator in the $i$-direction.

As shown in Ref.\cite{Benjamin}, at exactly the $K$-point($\bm{k} = \bm{0}$), the $\Omega^{inter}$-term in the second line of Eq.\ref{eq:Omegatot} arises from interband coupling between conduction and valence bands. It reduces to the orbital-type Berry curvature $\Omega_{orb}$. The $\Omega^{intra}$-term in the third line of Eq.\ref{eq:Omegatot} arises from intra-band coupling between the two conduction bands which are split by SOCs. It reduces to the spin-type Berry curvature $\Omega_{spin}$ at exactly the $K$-point. Thus, $\Omega_{tot} = \Omega_{spin} + \Omega_{orb}$ holds strictly at the $K$-point. It was also pointed out in Ref.\cite{Benjamin} that, based on the symmetry property of Berry curvature around $C_3$-invariant $K$-points, $\Omega_{tot} \approx \Omega_{spin} + \Omega_{orb}$ holds in the close vicinity of $K$.

Now, we show that throughout the Fermi level regime considered in this work, $\Omega^{intra} \approx \Omega_{spin}$ and  $\Omega^{inter} \approx \Omega_{orb}$ holds. As a result, we have $\Omega_{tot} \approx \Omega_{spin} + \Omega_{orb}$. To see why $\Omega^{intra} \approx \Omega_{spin}$, we note that at a finite momentum $\bm{k}$, the weight of the $\ket{d_{\textrm{v}}} \equiv \ket{d_{x^2-y^2} + i d_{xy}}$-states in the eigenstates $\ket{c,\pm, \bm{k}}$ of conduction bands is of the order $w_{\textrm{v}} \sim V^2_{F} k^2 /\Delta^2$ according to perturbation theory. As the relevant Fermi level ranges from $0 - 30$ meV measured from the conduction band minimum, the relevant momentum range covers $0 {\AA}^{-1}< k <0.06 {\AA}^{-1}$ measured from $K = 1.31 {\AA}^{-1}$(Fig.\ref{FIGS1}). With $V_F = 3.5 eV\cdot {\AA}$ and $\Delta = 1.66 eV$, the weight of $ \ket{d_{x^2-y^2} + i d_{xy}}$-state is $w_{\textrm{v}} < 2\%$ throughout this range. This suggests that in all ranges of Fermi level studied in this work, electronic states in the conduction band are strongly dominated by the $\ket{d_{\textrm{c}}} \equiv \ket{d_{z^2}}$-orbitals. 

The negligible contribution from $\ket{d_{x^2-y^2}+id_{xy}}$-states in the electronic wavefunctions allows us to construct an effective two-band Hamiltonian $H_{eff}(\bm{k})$ formed by $\{ \ket{d_{z^2},\uparrow}, \ket{d_{z^2},\downarrow} \}$-states for conduction band electrons within the range $0 {\AA}^{-1} < k < 0.06 {\AA}^{-1}$, such that the eigenstates $\ket{d_{z^2},\pm}$ with their corresponding eigenvalues $E_{\pm}(\bm{k})$ of $H_{eff}(\bm{k})$ are perturbatively good approximations for $\ket{c,\pm,\bm{k}}$ and $E_{c,\pm} (\bm{k})$ for evaluating $\Omega^{intra}$ in Eq.\ref{eq:Omegatot}. 

In fact, such an effective model can be obtained by treating intra-$\ket{d_{z^2}}$-orbital terms in $H_{tot}(\bm{k})$(Eq.\ref{eq:FourbandH}) as an unperturbed Hamiltonian, while the coupling between $\ket{d_{z^2}}$ and the remote $\ket{d_{x^2-y^2}+id_{xy}}$-states(which are predominantly in the valence band) in Eq.\ref{eq:FourbandH} are regarded as perturbations. Based on generalized second-order perturbation theory, we project contributions from the $\ket{d_{x^2-y^2}+id_{xy}}$-states onto the subspace of $\{ \ket{d_{z^2},\uparrow}, \ket{d_{z^2},\downarrow} \}$-states and obtain the effective two-band Hamiltonian (for $k \sim 0-0.06 {\AA}^{-1}$):
\begin{eqnarray}\label{eq:TwobandH}
H_{eff}(\bm{k}) &=& (\frac{\Delta V^2_F}{\Delta^2 - \lambda^2} k^2)\sigma_0 + \alpha_{so}\bm{k} \times \bm{\sigma}_{\parallel} \\\nonumber
&+& (\beta_0 + \beta'_1 k^2 + \frac{\lambda V^2_F}{\Delta^2 - \lambda^2} k^2) \sigma_z.
\end{eqnarray}

Here, the $\sigma$-Pauli matrices act on spins of $d_{z^2}$-orbitals, and $k^2 = k^2_x + k^2_y$. For simplicity, we drop all constant terms in Eq.\ref{eq:TwobandH} which have no contributions to Berry curvatures. Notably, the form of $H_{eff}$ in Eq.\ref{eq:TwobandH} is exactly the same as $H_0$(Eq.1 of the main text), which is derived based on the $C_{3v}$ point group symmetry. In particular, by defining $\beta_1 \equiv \beta'_1 + \lambda V^2_F/(\Delta^2 - \lambda^2)$, we identify the $\sigma_z$-term in $H_{eff}$ is exactly the Ising SOC term discussed in Eq.1 of the main text. Thus, the Berry curvature derived from $H_{eff}$, which serves as a perturbative approximation of $\Omega^{intra}$, is exactly the spin-type Berry curvature $\Omega_{spin}$ discussed in the main text. Therefore, we conclude that $\Omega^{intra}(\bm{k}) \approx \Omega_{spin}(\bm{k})$ throughout the regime explored in this work.

Next, we show that $\Omega^{inter} \approx \Omega_{orb}$ is also true for $k \sim 0-0.06 {\AA}^{-1}$. By the same reasoning above, it is straightforward to see that the weight of the $\ket{d_{z^2}}$-orbitals in the valence band states is also on the order of $V_F^2 k^2/\Delta^2 < 2\%$ within the range $k \sim 0-0.06 {\AA}^{-1}$. Thus, to evaluate the inter-band contribution $\Omega^{inter}$, the eigenstates $\ket{c,-,\bm{k}}$ and $\ket{v,\pm,\bm{k}}$ can be well approximated as $\ket{c,-,\bm{k}} \approx \ket{d_{z^2},-,\bm{k}}$ and $\ket{v,\pm,\bm{k}} \approx \ket{d_{x^2-y^2}+id_{xy},\pm,\bm{k}}$. We note that due to Rashba SOCs, the out-of-plane spin of eigenstates in the conduction band is no longer conserved. Thus, we have in general $\ket{d_{z^2},-,\bm{k}} = s_1(\bm{k}) \ket{d_{z^2},\uparrow} + s_2(\bm{k}) \ket{d_{z^2},\downarrow}$, and the $\bm{k}$-dependent coefficients $s_1, s_2$ need to be determined by diagonalizing $H_{eff}(\bm{k})$. However, by observing that $\Delta \gg \lambda \gg \beta_{0}, \beta_{1} k^2$ in the regime of our interest, we have $E_{c,-}(\bm{k}) - E_{v,\pm}(\bm{k}) \approx \Delta$. With $\ket{d_{x^2-y^2}+id_{xy},+(-),\bm{k}} = \ket{d_{x^2-y^2}+id_{xy},\uparrow(\downarrow),\bm{k}}$, we have:
\begin{widetext}
\begin{eqnarray}
\Omega^{inter}_{\textrm{c},-}(\bm{k}) &=& i \sum_{m=\pm}\frac{\braket{\textrm{c},-,\bm{k}|\hat{v}_x|\textrm{v},m,\bm{k}} \braket{\textrm{v},m,\bm{k}|\hat{v}_y|\textrm{c},-,\bm{k}} - c.c.} {(E_{\textrm{c},-}(\bm{k}) - E_{\textrm{v},m}(\bm{k}))^2} \\\nonumber
& \approx & -(|s_1(\bm{k})|^2 \frac{V_F^2} {\Delta^2} + |s_2(\bm{k})|^2 \frac{V_F^2} {\Delta^2}) \\\nonumber
& = & -\frac{V_F^2} {\Delta^2}
\end{eqnarray}
\end{widetext}
Note that in the last step we used the normalization condition $|s_1|^2+ |s_2|^2 =1$. As shown in Ref.\cite{Xiao}, in the limit $V_F k \ll \Delta$, $\Omega_{orb}$ is almost a constant with value $\frac{V_F^2} {\Delta^2}$ (a uniform momentum space profile in the neighborhood of $K$). It is clear that $V_F k \ll \Delta$ holds in the range  $k \sim 0-0.06 {\AA}^{-1}$ of our interest. Thus, we have $\Omega^{inter} \approx \Omega_{orb}$.

Based on our detailed analysis above, we conclude that $\Omega_{tot} \approx \Omega_{spin} + \Omega_{orb}$ holds throughout the regime explored in our work. We note that this relation would fail when $V_F k \sim \Delta$. However, this would require the Fermi level to exceed $100$ meVs\cite{Liu} which goes way beyond the regime of our interest. 

\begin{figure}
\centering
\includegraphics[width=3.5in]{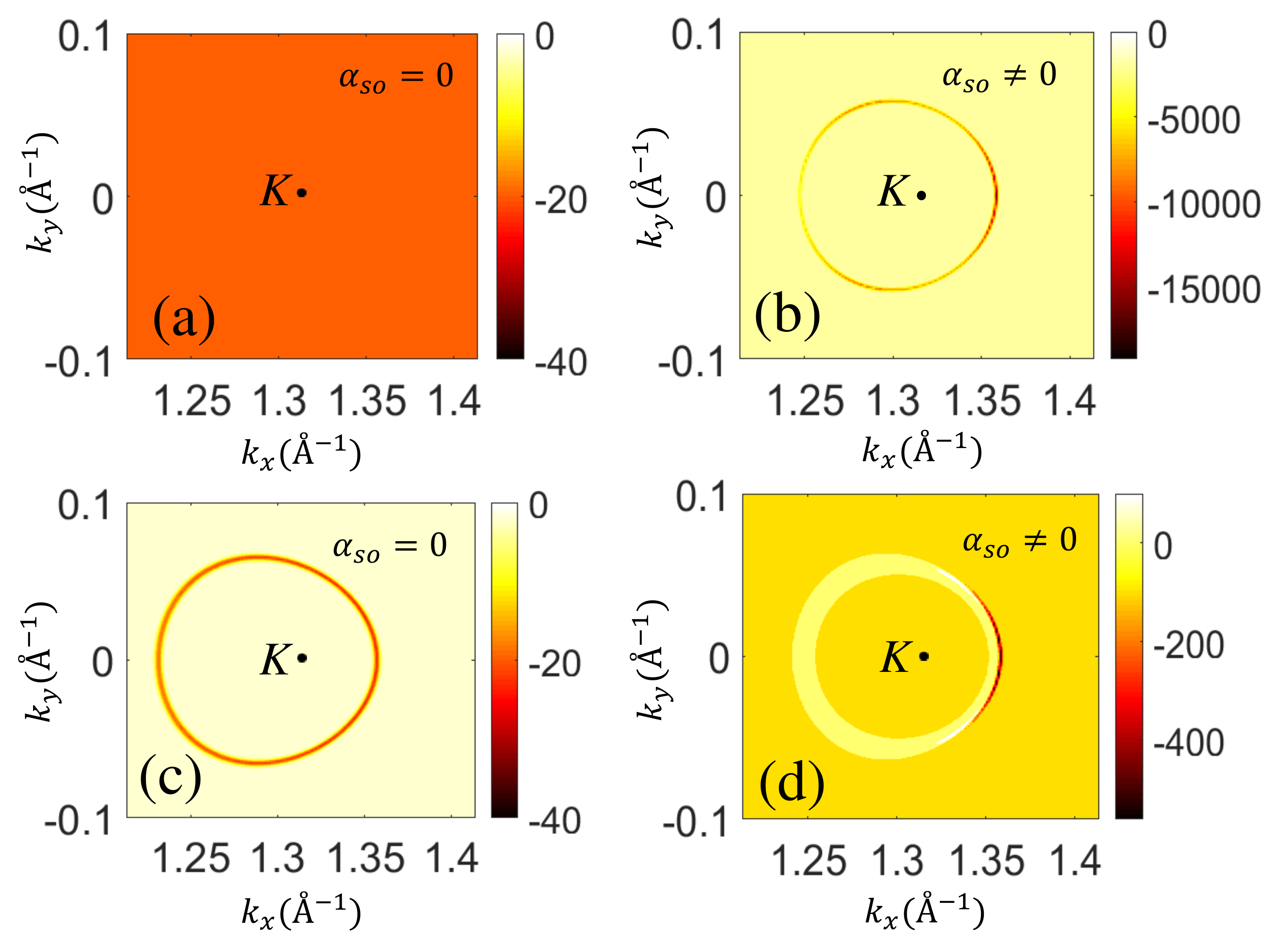}
\caption{Effects of $\Omega_{spin}$ on total Berry curvature profiles in strained MoSSe. Color bar indicates the value of total Berry curvature in units of ${\AA}^2$. (a)-(b) Profiles of $\Omega_{tot}$ of the lower conduction band $n=c,-$ throughout the neighborhood of $K$ in (a) strained MoS2 without $\Omega_{spin}$ and (b) strained MoSSe with $\Omega_{spin}$. (c)-(d) Profiles of $\Omega_{tot}$ along Fermi contour enclosing the $K$-point at $E_F =1.467$ eV in (c) strained MoS2 without $\Omega_{spin}$ and (d) strained MoSSe with $\Omega_{spin}$.}
\label{FIGS2}
\end{figure}

According to Fig.\ref{FIGS1}, $|\Omega_{spin}| \sim 10^4 {\AA}^2$ near the band anti-crossing points, while $|\Omega_{orb}| \sim 10 {\AA}^2$\cite{Xiao}. Thus, $\Omega_{spin}$ strongly dominates over $\Omega_{orb}$ near the band anti-crossing points. Based on our discussions in subsection A, the relation $\Omega_{tot} \approx \Omega_{spin} + \Omega_{orb}$ implies that the behavior of $\Omega_{tot}$ is essentially governed by $\Omega_{spin}$ in the regime of our interest. In this section, we explicitly demonstrate how the dominance of $\Omega_{spin}$ gives rise to strongly enhanced Berry dipole of order $1 {\AA}$ near the band anti-crossing points as shown in Fig.2-3 of the main text.

First, to demonstrate how $\Omega_{spin}$ significantly enhances the total Berry curvature in strained MoSSe, we plot the profiles of $\Omega_{tot}$ of the lower conduction band $n=c,-$ \textit{throughout the neighborhood of $K$} as shown in Fig.\ref{FIGS2}a-b. In Fig.\ref{FIGS2}a, the Rashba SOC is absent($\alpha_{so}=0$), thus $\Omega_{spin}=0$ according to Eq.2 of the main text. In this case, the magnitude of $\Omega_{tot}$ remains almost a constant at $\sim 10 {\AA}^2$. Notably, as we discussed in subsection A, the profile of $\Omega_{tot}$ is approximately a uniform function near $K$ due to the fact that $V_F k \ll \Delta$\cite{Xiao}. 

In contrast, when Rashba SOC is present, the maximum amplitude of $\Omega_{tot}$ is significantly enhanced near the band anti-crossing points due to the presence of $\Omega_{spin}$(Fig.\ref{FIGS2}b). Note that the magnitude of $\Omega_{tot}$ is also on the order of $10^4 {\AA}^2$, being consistent with the relation: $\Omega_{tot} \approx \Omega_{spin} + \Omega_{orb}$. Importantly, as the magnitude of $\Omega_{spin}$ changes dramatically as $\bm{k}$ goes from $K$ to the band anti-crossing points(Eq.2 of the main text), the profile of $\Omega_{tot}$ also becomes highly nonuniform in the neighborhood of $K$ as explicitly shown in Fig.\ref{FIGS2}b. 

Second, to demonstrate how $\Omega_{spin}$ leads to strongly enhanced Berry dipole in strained MoSSe, we plot the Berry curvature profiles \textit{along a fixed Fermi contour enclosing the $K$-point}(i.e., only momentum states at a given Fermi energy $E_F$ are considered) as shown in Fig.\ref{FIGS2}c-d. This provides a physical picture of a Berry dipole according to Eq.4 of the main text, which can be regarded as a manifestation of its Fermi liquid property. As shown in Fig.\ref{FIGS2}c, when $\Omega_{spin}$ is absent, the total Berry curvature $\Omega_{tot}$ is also approximately a constant along the Fermi contour due to the uniform Berry curvature profile throughout the neighborhood of $K$(Fig.\ref{FIGS2}). As the Berry dipole is given by the sum of products between Fermi velocity and Berry curvature at each $\bm{k}$ along the Fermi surface(Eq.4 of the main text), the contributions from all states $\bm{k}$ in Fig.\ref{FIGS2}c almost cancel each other because their Berry curvatures are roughly the same while their velocities almost sum to zero.

In sharp contrast, with the same Fermi contour in Fig.\ref{FIGS2}c, due to the highly nonuniform profile of $\Omega_{tot}$ in Fig.\ref{FIGS2}b, $\Omega_{tot}$ also changes drastically along the Fermi contour as shown in Fig.\ref{FIGS2}d. Note that due to the broken $C_3$ symmetry by strain, the Berry curvature profile is no longer three-fold invariant about $K$. Particularly, with the choice of Fermi energy lying close to hot spots of $\Omega_{spin}$ associated with the right-movers, the Berry curvature along the Fermi contour enclosing $K$ is concentrated preferentially on the right hand side of $K$(Fig.\ref{FIGS2}d). In this case, the largely imbalanced Berry curvature contributions from right-movers and left-movers lead to a large Berry dipole according to Eq.4 of the main text.

\begin{figure}
\centering
\includegraphics[width=3.5in]{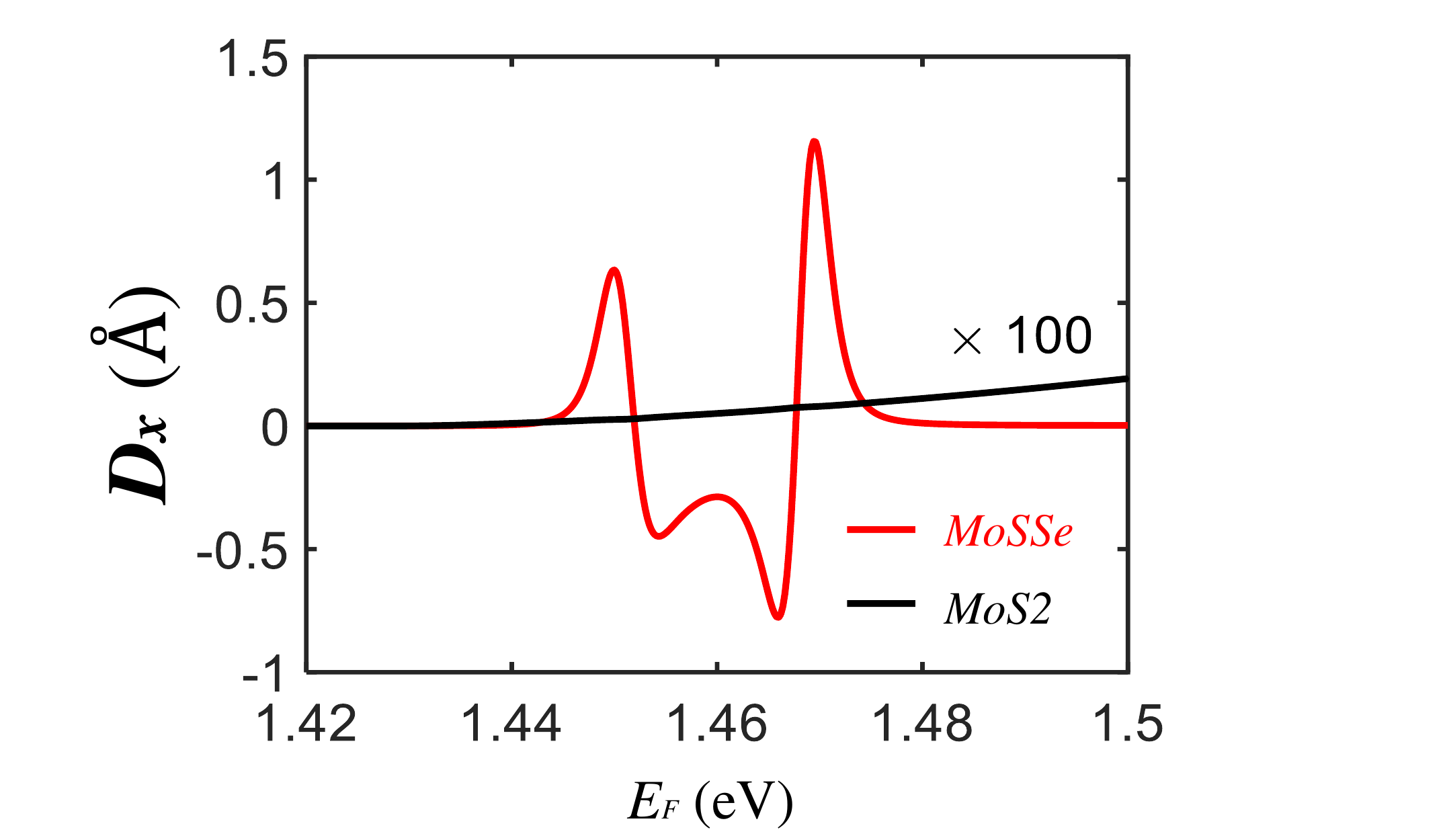}
\caption{Gate-dependence of Berry curvature dipole $D_x$ in strained MoS2 (black solid line) and strained MoSSe (red solid line) with $u_{xx} = 4\%$. Note that $D_x$ in strained MoS$_2$ is multiplied by 100 times so that the values can be visible on the scale of $0.5 {\AA}$. For direct comparison, all parameters used for calculating both curves are set to be the same except that $\alpha_{so} \neq 0$ for MoSSe. }
\label{FIGS3}
\end{figure}

We now explicitly demonstrate the enhancement of Berry dipole due to $\Omega_{spin}$. We calculate the Berry dipole for strained conventional MoS$_2$ without Rashba SOCs for a direct comparison with strained MoSSe with Rashba SOCs. As shown clearly in Fig.\ref{FIGS3}, the magnitude of $D_x$ in MoS$_2$ is only of the order $1 \times 10^{-3} {\AA}$ in the regime where $D_x$ is of order $1 {\AA}$ in strained MoSSe. Thus, $D_x$ is strongly enhanced by 3 orders of magnitude in the Fermi level regime $E_F \sim 1.44 - 1.48$ eV due to the presence of $\Omega_{spin}$. We note by passing that the optimal value for $D_x$ in strained MoS$_2$ (with contributions from $\Omega_{orb}$ only) can be as large as $\sim 0.01 {\AA}$ at a higher $E_F$ which goes beyond the regime considered here. Thus, the optimal value is also enhanced by 2 orders of magnitude due to $\Omega_{spin}$.

\renewcommand{\theequation}{E-\arabic{equation}}
\renewcommand\thefigure{E-\arabic{figure}}  
\setcounter{equation}{0}  
\setcounter{figure}{0}
\section*{\textbf{Appendix E: Berry curvature dipole in other strained polar TMDs}}

\begin{figure}
\centering
\includegraphics[width=3.5in]{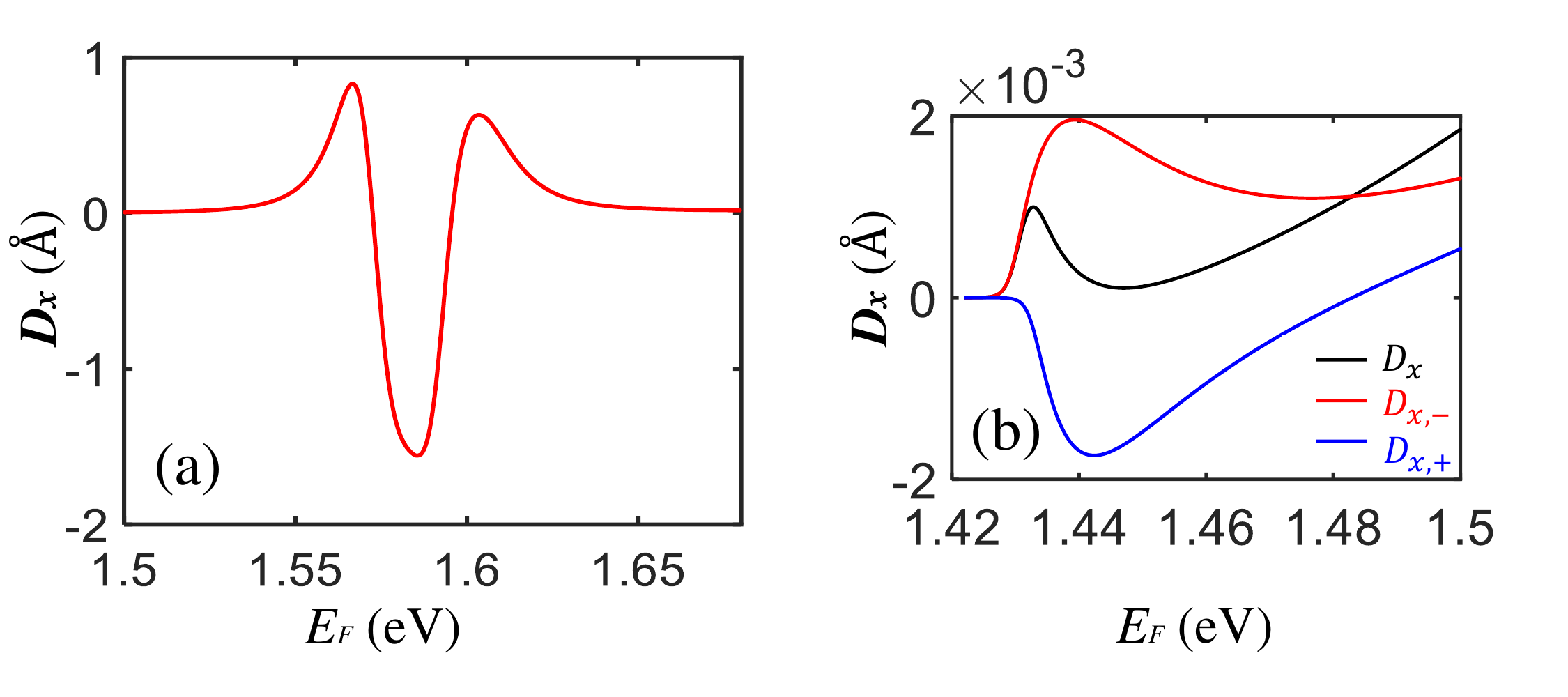}
\caption{Gate-dependence of Berry curvature dipole $D_x$ in (a) strained MoSeTe with $u_{xx} = 2\%$ and (b) strained WSeTe with $u_{xx} = 4\%$. }
\label{FIGS4}
\end{figure}

Here, we discuss nonlinear Hall effects in other strained polar TMDs. First, as we mentioned in the main text, our prediction of large Berry dipole generally applies to the whole class of moly-based polar TMDs. This is due to the fact that in all conventional moly-based TMDs, there readily exists a band crossing within the two electron bands\cite{Liu}. As a result, once Rashba SOCs are turned on, the band crossing points are gapped out, leading to band-anticrossings where Berry curvature hot spots can emerge. 

To explicitly demonstrate the generality of the mechanism above, we calculate the gate-dependence of Berry dipole of another moly-based TMD material, MoSeTe, as shown in Fig.\ref{FIGS4}a under uniaxial strain $u_{xx} = 2\%$. The tight-binding model used for MoSeTe is exactly the same as the one presented in Section II and III, with the parameters for MoSeTe listed in Table III and IV. The SOC tight-binding parameters for MoSeTe are given by $\alpha_0 = 1.2$ meV, and $\beta_{so} = -15$ meV. Clearly, $D_x$ in MoSeTe has the same qualitative gate-dependent features as MoSSe under similar strain $u_{xx} \sim 2\%$(Fig.3 of the main text). Notably, the optimal value of $D_x$ in MoSeTe is also of the order $1 {\AA}$.

On the other hand, as we pointed out in the main text, in tungsten(W)-based 2H-TMDs, the SOC-induced Berry curvature $\Omega_{spin}$ can also modify the total Berry curvature profile near the conduction band minimum. However, since there is no band anti-crossing in W-based materials, the Berry curvature profile is much less nonuniform compared to the moly-based case and the Berry curvature dipole is expected to be less strongly enhanced. Here, using a specific example of W-based polar TMD candidate WSeTe, we demonstrate this fact by calculating the Berry curvature dipole under the same uniaxial strain $u_{xx} = 4\%$ as in Fig.2 of the main text. For W-based materials, the Ising SOC in the conduction band is set to be $\beta_0 \approx 20$ meV\cite{Liu}. Strong Rashba splitting $\sim 50$ meV has also been predicted for WSeTe\cite{Wan}, and the fitted parameter in the tight-binding model is set to be $\alpha_{0} \approx 45$ meV according to previous works\cite{Benjamin}. For simplicity, the other parameters are set to be the same as those presented in Table \ref{table:02} and Table \ref{table:03}.

The Berry curvature dipole $D_x$ as a function of the Fermi energy $E_F$ for strained WSeTe is shown in Fig.\ref{FIGS4}b. Clearly, the Berry curvature dipole is also on the order of $1 \times 10^{-3}$ ${\AA}$, comparable to the case of conventional 2H-TMDs while much smaller than the values in strained MoSSe (Fig.2 of the main text). However, due to the fact that $\Omega_{spin}$ causes the two bands to carry opposite Berry curvatures, the contributions $D_{x,-}/D_{x,+}$ from lower/upper bands to $D_x$ also have different signs as shown in Fig.\ref{FIGS4}b, which results in some cancellation effects as the upper band gets filled (signified by the point in the $D_{x} - E_F$ curves where $D_{x}$ starts to deviate from $D_{x,-}$). While certain non-trivial gate-dependence in $D_x$ is also found, the sign of $D_x$ is not switched as in the case of MoSSe (Fig.2 $\&$ 3 of the main text).

\end{document}